\documentclass{article}

\usepackage{PRIMEarxiv}

\usepackage[utf8]{inputenc} % allow utf-8 input
\usepackage[T1]{fontenc}    % use 8-bit T1 fonts
\usepackage{hyperref}       % hyperlinks
\usepackage{url}            % simple URL typesetting
\usepackage{booktabs}       % professional-quality tables
\usepackage{amsfonts}       % blackboard math symbols
\usepackage{amsmath}
\usepackage{nicefrac}       % compact symbols for 1/2, etc.
\usepackage{microtype}      % microtypography
\usepackage{lipsum}
\usepackage{graphicx}
\usepackage{subcaption}
\usepackage{mathpazo}
\usepackage{sectsty}
\usepackage{tikz}
\usepackage[capitalize]{cleveref}
\usepackage[shortlabels,inline]{enumitem}
\usepackage{float}
\usepackage{natbib}
\usepackage{csquotes}
\usepackage{tabularx}
\newcolumntype{L}{@{\extracolsep{\fill}}l}
\newcolumntype{R}{@{\extracolsep{\fill}}r}
\newcolumntype{C}{@{\extracolsep{\fill}}c}
\usepackage{placeins}
\usepackage{ltablex}
\keepXColumns
\usepackage[ruled,vlined]{algorithm2e}
\usepackage{subcaption}
\usepackage[most]{tcolorbox}\usepackage{enumitem}
\newtcolorbox[auto counter, number within=section]{algobox}[2][]{%
  float,                        % lo rende float
  floatplacement=ht!,           
  colback=white, colframe=black!40,
  title=Box~\thetcbcounter: #2,
  #1
}
\title{System Identification of a Moored ASV with Recessed Moon Pool via Deterministic and Bayesian Hankel-DMDc}

\author{
  Giorgio Palma$^{a, \star}$, Ivan Santic$^{a}$, Andrea Serani$^{a}$, Lorenzo Minno$^{b}$, Matteo Diez$^{a}$\\
  $^{a}$National Research Council-Institute of Marine Engineering, Via di Vallerano 139, Rome, 00128, Italy\\
  $^{b}$Codevintec Italiana S.r.l., Viale Lenormant 215-217, Rome, 00119, Italy\\
  $^\star$\texttt{giorgio.palma@cnr.it} \\
}

\begin{document}

\begin{tikzpicture}[remember picture,overlay]
   % Nodo per il riempimento con trasparenza
   \node [rectangle, fill=cyan, fill opacity=0.5, anchor=north, minimum width=\paperwidth, minimum height=3cm] at (current page.north) {};

   % Nodo separato per il testo, senza trasparenza
   \node [anchor=north, minimum width=\paperwidth, minimum height=3cm, text width=\textwidth, align=center, text height=5ex, text depth=15ex, align=left] at (current page.north) {
     \sffamily\small
     \textbf{This is a preprint submitted to:} \textit{Journal of Marine Science and Engineering}
   };
\end{tikzpicture}

\maketitle

\begin{abstract}
This study addresses the system identification of a small autonomous surface vehicle (ASV) under moored conditions using Hankel dynamic mode decomposition with control (HDMDc) and its Bayesian extension (BHDMDc). Experiments were carried out on a Codevintec CK-14e ASV in the towing tank of CNR-INM, under both irregular and regular head-sea wave conditions. The ASV under investigation features a recessed moon pool, which induces nonlinear responses due to sloshing, thereby increasing the modelling challenge. Data-driven reduced-order models were built from measurements of vessel motions and mooring loads. The HDMDc framework provided accurate deterministic predictions of vessel dynamics, while the Bayesian formulation enabled uncertainty-aware characterization of the model response by accounting for variability in hyperparameter selection. 
Validation against experimental data demonstrated that both HDMDc and BHDMDc can predict the vessel's response to unseen regular and irregular wave excitations. 
{\color{black}In conclusion, the study shows that HDMDc-based ROMs are a viable data-driven alternative for system identification, demonstrating for the first time their generalization capability for a sea condition different from the training set, achieving high accuracy in reproducing vessel dynamics.}
\end{abstract}

% keywords can be removed

%%%%%%%%%%%%%%%%%%%%%%%%%%%%%%%%%%%%%%%%%%
\section{Introduction}
The maritime and ocean engineering communities have increasingly recognized the importance of developing accurate and computationally efficient models for the prediction and control of marine system dynamics. 
In particular, the capability to accurately predict ship motions is essential for supporting their design, monitoring, and decision-making tasks during their operations.
This holds particularly for seakeeping and maneuvering in adverse weather conditions, for which the availability of reliable predictive tools helps in the development and operation of vessels, ensuring the safety of structures, payload, and crew.
In this regard, commercial and military ships must meet the International Maritime Organization (IMO) Guidelines and NATO Standardization Agreements (STANAG), respectively. 
Both regulatory frameworks emphasize the need for accurate assessments of vessel motions and loads under a wide range of sea conditions, highlighting the importance of robust modeling and simulation tools to support compliance and operational readiness.

The complexity and nonlinear nature of hydrodynamic phenomena involved pose significant challenges to predicting ship responses in waves.
Recent works \citep{stern2015, serani2021urans, aram2024cfd} demonstrated the ability of computational fluid dynamics (CFD) methods with unsteady Reynolds-averaged Navier-Stokes (URANS) formulations to assess ship performances in waves and extreme sea conditions.
% Other studies employed potential flow solvers \citep{lin2006numerical,BelknapReed2019}, and also hybrid formulations where URANS and potential flow are combined to reduce computational costs and yet model important viscous-related features \citep{White2022}.
Along with the high fidelity of simulations, however, comes their high computational costs. 
This is particularly true when simulations aim to achieve statistical convergence of relevant quantities of interest, and complex fluid-structure interactions are investigated.
Real-time applications, such as control, fault detection, and digital twinning, are also limited by the computational effort required from such models.

In this context, data-driven modeling techniques have emerged as powerful alternatives or complements to traditional first-principles approaches. 
They promise to reduce the computational cost while keeping the fidelity of their estimates comparable to the original data sources, given that they are properly trained and/or calibrated, which is generally not trivial.
Within this framework, system identification provides a structured approach for constructing predictive models able to incorporate key characteristics of the system  from high-fidelity numerical tools and experiments. Here, we interpret system identification broadly as the development of reduced-order models (ROMs) capable of predicting system responses, independent of whether their parameters correspond directly to physical quantities.

Equation-based reduced order models (ROMs), such as the Maneuvering Modeling Group (MMG) model \cite{Yasukawa2015,Yasukawa2016}, have been developed as physics-based efficient approaches, and demonstrated good agreement with experiments and CFD for maneuvering of displacement ships, \cite{Sanada2021}, twin-hull configurations \cite{pandey2016manoeuvring,PANDEY2016}, and have been recently studied also for planing hulls \cite{diez2024sname}. 
Valuable characteristics of physics-based models are their fast evaluation and the possibility of gaining knowledge about the analyzed phenomena through the interpretation of the solution and the identified system, typical of methods classifiable as white-box (fully equation-based) or grey-box (physics-based with data tuning).
Despite the promising results, such models typically exhibit a limited adaptability to unmodelled dynamics (white-box) or require a large amount of data from CFD computations or EFD for their training and definition of forcing terms (grey-box).

Data-driven machine learning techniques gained popularity due to their ability to model complex input-output relations in an automated manner directly from data, not requiring the effort of gaining prior knowledge on the system to develop a full model. 
In particular, recurrent neural networks (RNN) \citep{DAgostino2022} and long short-term memory networks (LSTM) \citep{XU2021}, along with their bidirectional LSTM (BiLSTM) \citep{Wang2023, jiang2024} variant, have been demonstrated to be effective for building equation-free data-driven models for ship motions, and provide multi-step ahead forecasting of the ship's degree-of-freedom in several sailing conditions, including calm water and waves \citep{Diez2024}.
The strength of machine and deep learning methods lies in their ability to capture relevant hidden and nonlinear dynamics and input-output relations directly from available data, their compactness, and fast evaluation. 
However, deep learning models typically require large datasets of high-fidelity data for training (more complex architectures usually require more expensive training) and to generalize effectively.
In addition, while powerful, such models are often considered black-box approaches and pose challenges in terms of the physical interpretability of their results.
In addition,

Among the available methods, the dynamic mode decomposition (DMD) \citep{schmid2010,kutz2016dynamic,mezic2021koopman, Brunton2021} and its methodological variants have recently gained attention due to their ability to extract dominant dynamic features directly from experimental or numerical data, with small or no assumption on the underlying physics, providing interpretable, low-dimensional representations of nonlinear systems. 
DMD can be classified somewhere between a black-box and a grey-box approach for reduced-order modeling. With the former, it shares the data-driven and equation-free structure like other machine learning techniques, such as RNN, LSTM, and BiLSTM. However, DMD-based methods retain a certain level of interpretability thanks to the linear nature of the model.
DMD can be considered a method to build a finite-dimensional approximation of the Koopman operator \citep{Koopman1931}, which, in turn, describes a nonlinear dynamical system as a possibly infinite-dimensional linear system \citep{Proctor2018}. 
The reduced-order linear model is obtained by DMD from a small set of multidimensional snapshots of the dynamical system under analysis. The DMD operates equally on measured or simulated data and obtains the model from them with a direct procedure that, from a machine learning perspective, constitutes the training phase. 
Its data-driven nature, the fast, non-iterative training, and data-lean property contributed to the popularity of DMD as a reduced-order modeling technique in several fields, such as fluid dynamics and aeroacoustics \citep{rowley2009, schmid2010, Tang2012, Semeraro2012, Song2013}, epidemiology \citep{Proctor2015}, neuroscience \citep{brunton2016}, finance \citep{mann2016}, etc.
DMD was applied for the first time to the forecasting of ship dynamics in \cite{diez2022datadriven}, in which the proof of concept of short-term forecasting of trajectories, motions, and loads of maneuvering ships in waves was given.
In \cite{serani2023}, the approach was systematically assessed on the same test cases and first extended to the use of Augmented-DMD by augmenting the system state with lagged copies of the original states and their derivatives. This approach enabled the modeling of memory effects in the system, improving accuracy over the tested cases compared to the standard formulation.
\citep{palma2024forecasting} systematically explored the use of Hankel-DMD (HDMD) for short-term forecasting of ship motions, highlighting its potential for real-time prediction and control applications, and digital twinning.
The first effort into the development of a DMD-based data-driven system identification method to build an input/output reduced order model for ship motions was conducted in \cite{Serani2024snh}. A model predicting the roll motion of a seakeeping vessel from the knowledge of the wave elevation was obtained by using the DMD methodological extension called dynamic mode decomposition with control (DMDc). This incorporates the control variables and forcing inputs in the system regression, separating their effect from the free evolution of the system. 
Input inclusion and Hankel extension have been combined in Hankel-DMD with control (HDMDc) and applied for the first time to ship motion and forces prediction in \cite{palma2025si}, which demonstrated the capability of the method to achieve good accuracies without degradation through the observation time.
The HDMDc was then tested on several applications, introducing methodological advancements to face the challenges of the specific cases, such as the embedding of nonlinear observables in the state and input vectors to address extremely nonlinear responses of planing hulls in slamming \citep{palmaFAST2025}, and the use of a Tikhonov-regularized least-square formulation for improving the numerical stability of the DMD regression when using noisy experimental data \citep{palma2025sicatamaran}.

Quantifying the uncertainty associated with the prediction of a ROM has become increasingly relevant for data-driven modeling, and a key characteristic for their usage in the context of, \textit{e.g.,} multifidelity analysis and optimization.
Few approaches for introducing uncertainty quantification in DMD analysis have been presented in the literature so far. \cite{takeishi2017b} first introduced a probabilistic model by modeling the measurement noise as Gaussian and treating the dynamic modes and eigenvalues as random variables, whose posterior distributions were inferred through Gibbs sampling. 
Later, \cite{Sashidhar2022} presented the bagging optimized-DMD, where Breiman’s statistical resampling strategy was applied to training data, generating ensembles of DMD models and estimating confidence intervals for the extracted modes and eigenvalues.
A similar approach was used in \cite{palma2025sicatamaran} and called frequentist approach: several training signals were used, producing an ensemble of HDMDc models and estimating confidence intervals for time-resolved predictions of ship motions.
\cite{Serani2024snh} considered for the first time the uncertainty arising from the selection of HDMD and HDMDc hyperparameters: the length of the training sequence and the number of time-lagged copies of the state were separately considered as probabilistic variables with uniform distributions within a suitable range. Monte Carlo sampling is applied to obtain an ensemble of predictions forming a posterior distribution.
This concept, referred to as Bayesian, has been further developed in \citep{palma2024forecasting, palma2025si, palma2025sicatamaran, palmaFAST2025}, where all the method hyperparameters are considered stochastic variables at once.
%The Bayesian formulation defines a prior probability density for the hyperparameters of the Hankel-DMD, obtaining a posterior of the predicted variables. We note that the proposed method is termed “Bayesian” in a broader sense than classical Bayesian inference. In particular, posterior distributions are not inferred over the Koopman operator or dynamic mode decomposition matrices themselves. Instead, a Bayesian treatment of the Hankel-DMD hyperparameters (observation window length and delay embedding) is adopted, assigning them uniform prior distributions and propagating their uncertainty through Monte Carlo sampling. This results in a predictive distribution for the ship motion time series, characterized by mean and standard deviation. While this does not constitute full Bayesian parameter inference, it is consistent with Bayesian principles of uncertainty propagation and model averaging, and aligns with how “Bayesian” is used in related literature for uncertainty-aware reduced-order modeling.

HDMDc and its stochastic extensions enabled the construction of robust, uncertainty-aware ROMs capable of capturing the essential dynamics of marine systems under realistic operating conditions.
In practical marine applications, it is crucial that ROMs remain valid across a variety of sea states, wave spectra, and loading conditions. Testing the transferability of DMD-derived models beyond their training datasets therefore represents a key step toward their reliable adoption in real-world scenarios.
However, despite the growing body of literature on DMD-based modeling in ship dynamics, most studies so far have focused on fixed or well-controlled experimental conditions, with limited exploration of the generalization capabilities of such models when exposed to new or unseen environments. 
In \citep{palma2025si, palma2025sicatamaran, palmaFAST2025}, for example, the test set for the DMD-based ROMs was composed from ship's dynamic responses to new and unseen forcing wave signals, which, however, represented different realizations of the same sea state, characterized by an identical spectral distribution of wave energy. Consequently, although the wave sequences used for testing were unseen during training, they were statistically consistent with the training conditions, thus not probing the generalization capability of the models to different sea states. 

The present work aims to address this gap by investigating the capability of HDMDc and BHDMDc system identification to generalize beyond the training conditions.
A dedicated experimental campaign was performed at the CNR-INM towing tank facility, collecting data from a Codevintec CK-14e autonomous surface vehicle (ASV) subjected to irregular and regular head-sea wave conditions. The ASV features a recessed moon pool, which induces significant nonlinearities in the hydrodynamic response, primarily associated with sloshing and piston-like oscillations of the internal free surface. 
Specifically, the DMD-based ROMs are learned using data from the irregular waves condition and subsequently applied to predict the experimental vessel response also in regular wave conditions. 
Statistical and probabilistic analyses were employed to quantify prediction accuracy and uncertainty.

The remainder of this paper is organized as follows. \Cref{s:exp} describes the experimental setup and data acquisition. \Cref{s:dmd} presents the HDMDc and BHDMDc methodologies adopted in this study. \Cref{s:res} discusses the identification results and the validation of the models under different sea conditions. Finally, \cref{s:conc} summarizes the main findings and outlines perspectives for future research on data-driven modeling of marine systems.

%%%%%%%%%%%%%%%%%%%%%%%%%%%%%%%%%%%%%%%%%%
\section{Experimental setup} \label{s:exp}
\subsection{Facility and model}
Experimental tests were carried out at the CNR-INM Emilio Castagneto seakeeping basin in Rome. The facility is 220 m long, 9 m wide, and 3.6 m deep. It is equipped with a single-flap wave generator capable of producing both regular and irregular waves: regular waves with wavelengths ranging from 1 to 10 m and corresponding heights from 0.1 to 0.45 m, as well as irregular waves following any desired sea spectrum at the appropriate scale. The plunger flap, whose rotation axis is located at 1.80~m below the calm water level (mid-depth), is electro-hydraulically driven by three pumps with a total power of 38.5 kW. Its deflection angle is controlled in the range $\pm 13$~deg through an electronic programming system producing harmonic components, each modulated in both amplitude and frequency (in the range 0.1-1.4~Hz).

A full-scale Codevintec CK-14e autonomous surface vessel was used for the present work. The vessel is a small marine drone, with a carbon fiber and Kevlar hull, 1.40~m in length ($L_{m}$), 0.9~m in beam, and 0.35~m in height (0.45~m including the rollbar). In the configuration used for the present tests, the model weighed 60.7~kg, including 22~kg from the hull, 13.4~kg from the batteries (mounted to replicate the actual free-running setup of the vessel), and the remainder from payload (\textit{i.e.}, cables and instrumentation). 
%Even though the model was moored for the test, batteries were mounted to replicate the actual free-running setup for the vessel.

\begin{figure}
    \centering
    \captionsetup[subfigure]{justification=centering}    
    \begin{subfigure}[b]{0.354\linewidth} 
       \includegraphics[width=\linewidth]{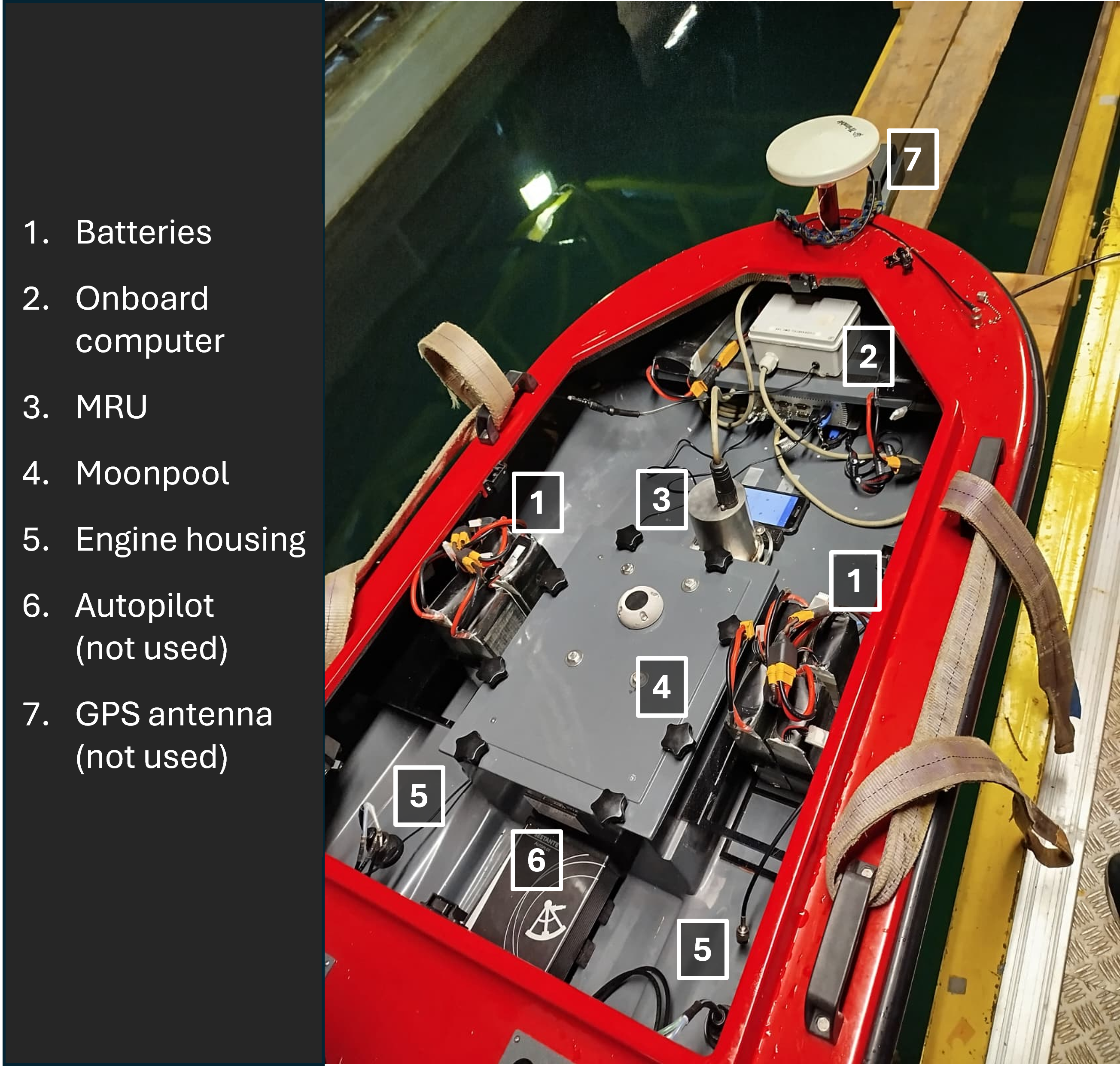}
        \caption{}   \label{fig:ck14}
    \end{subfigure}
    \begin{subfigure}[b]{0.6\linewidth} 
        \includegraphics[width=\linewidth]{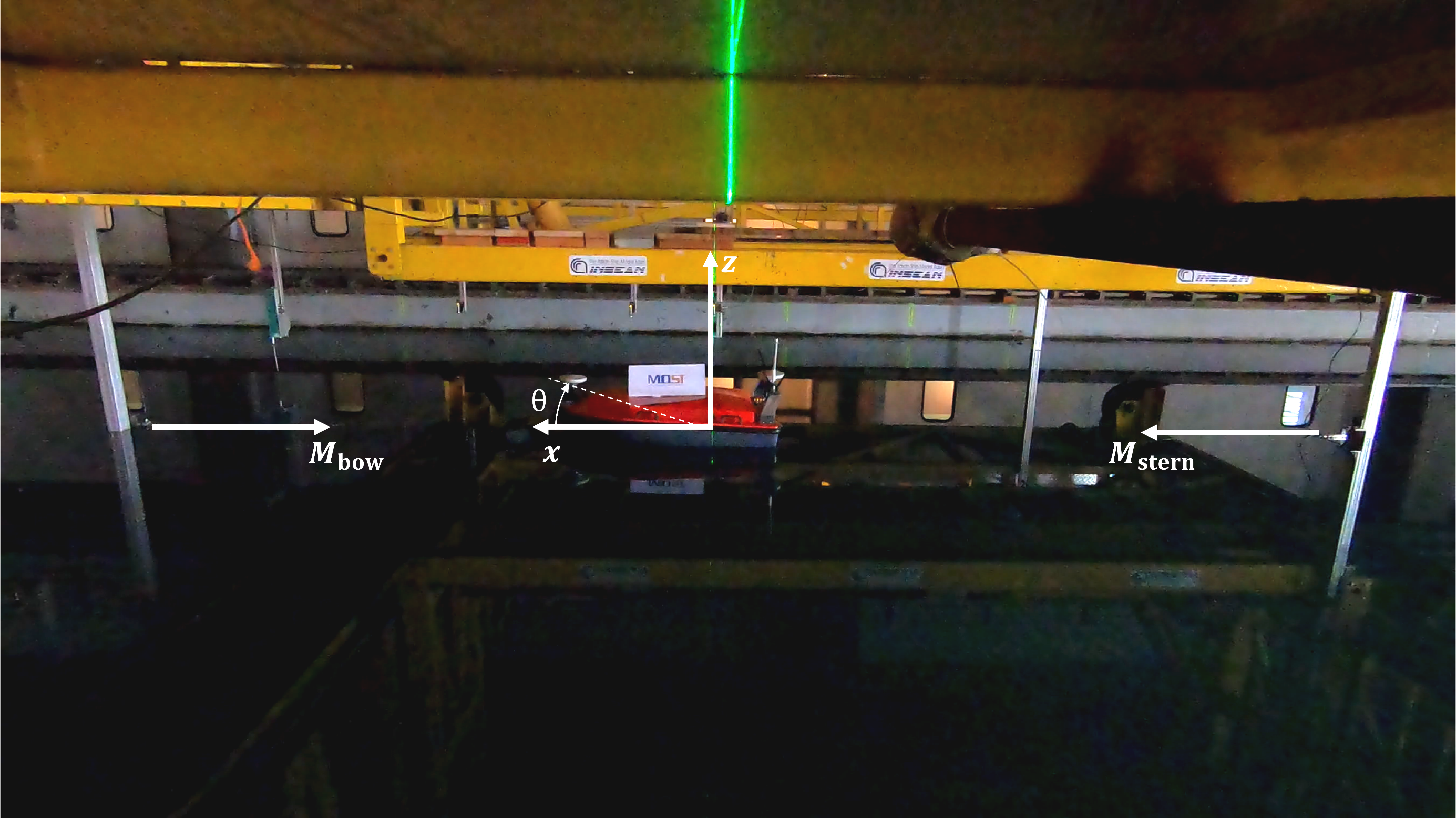}
        \caption{}   \label{fig:testfield}
    \end{subfigure}
    \caption{Picture of the CK-14e ASV (\subref{fig:ck14}), the top cover was removed, showing the internal configuration and the location of the moon pool, visible from the lid and its fastening system. Towing tank test rig for CK-14e in moored conditions, the white panel used in tracking is visible on the model top cover (\subref{fig:testfield}).}  
\end{figure}

\subsection{Moored tests}
During the tests, the CK-14e model was moored in the towing tank and encountered head irregular and regular waves, as shown in \cref{tab:testwaves} and described below.

A stationary carriage in the tank was used to position the bow and stern mooring poles for the model; the distance between the two poles was 8.22~m, 
%(the actual distance between mooring points was 7.85~m)
with the bow pole positioned 48.24~m away from the wave generator flap.
The model was moored to the poles through elastic mooring lines, characterized by linear elastic constants $k_{\mathrm{bow}}=1.89$~N/m and $k_{\mathrm{stern}}=1.33$~N/m.
 
For both regular and irregular wave testing, sea state conditions were selected by considering the CK-14e as a 1:50 scaled model of a 70~m long supply vessel.
% Given the characteristics of the facility, developed to perform tests with a scaled model of ships, sea state conditions have been selected hypothesizing a scale factor of 50 (TBC), thus considering the drone being a 70m long supply vessel.
%
\begin{table}[ht!]
    \centering
\caption{Test matrix, regular and irregular wave conditions}\label{tab:testwaves}
    \begin{tabular}{lllll}
\toprule
    $h_s$ [m]    & $\lambda_w/L_{m}$ [-]  & $v_{phase}$ [m/s] & $v_{group}$ [m/s] & performed tests\\
\midrule
    0.15 & \multicolumn{3}{l}{Pierson-Moskowitz spectrum} & 3\\
    0.10 & 1.5 &1.811 &0.905 & 1\\
    0.10 & 2   &2.090 &1.045 & 1\\
    0.10 & 3   &2.560 &1.280 & 1\\
    \midrule
    \end{tabular}
\end{table}
\subsubsection{Irregular waves}
The first part of the experiments was dedicated to model testing in head irregular waves.
{\color{black}The wave generator was operated to produce a linear superposition of $N_w=100$ elemental wave components with random phase $\varphi$ and frequencies equally spaced in $ 0.28$~Hz$\le f\le 1.13$~Hz
\begin{equation} \label{eq:irr_wave}
\eta(t) = \sum_{i=1}^{N_w} A(f_i)\sin(2\pi f_it+\varphi_i).
\end{equation}
The amplitudes $A(f_i)=\sqrt{(4\pi E(f_i)\Delta f_i)}$. 
were defined to obtain a the sea state 7 Pierson-Moskowitz (P-M) spectrum $E(f)$ \cite{carter1982} characterized by a significant height of $h_s=0.15$~m and peak period $T_p=2.12$~s in model scale ($h_s=7.5$~m and $T_p=15$~s in full scale):
\begin{equation}\label{eq:pm}
E(f) = \frac{5}{16}h_s^2 T_p^{-4}f^{-5}\exp\left(-\frac{5}{4}T_p^{-4}f^{-4}\right).    
\end{equation}
% \begin{equation}
%     T_z = \sqrt{\frac{\int E(f)\, df }{\int f^2 E(f) \,df}}.
% \end{equation}
\Cref{fig:pmspectrum} compares the nominal generated power spectral density from \cref{eq:pm} with the corresponding from the experimental wave elevation measured by the Kenek wave elevation probe in the forward position (see \cref{tab:waveprobes}). The EFD power spectral density is obtained using the Welch's method by analyzing approximately 210~s of wave elevation signal per run (10.5 minutes) in model scale, corresponding to about 4455~s in full scale (1 hour and 14 minutes). The signal, sampled at $500~Hz$, was divided into segments of $2^{12}$ samples with an overlap of $2^7$ samples. Each segment was transformed using a $2^{16}$-point FFT, and the resulting periodograms were averaged to obtain the final spectral estimate. 
The actual realizations are characterized by an average significant height of $\hat h_s=0.16$~m and average peak period of $\hat T_p=1.98$~s in model scale ($\hat h_s=8$~m and $\hat T_p=14$~s in full scale).
\begin{figure}
    \centering
    \includegraphics[width=0.75\linewidth]{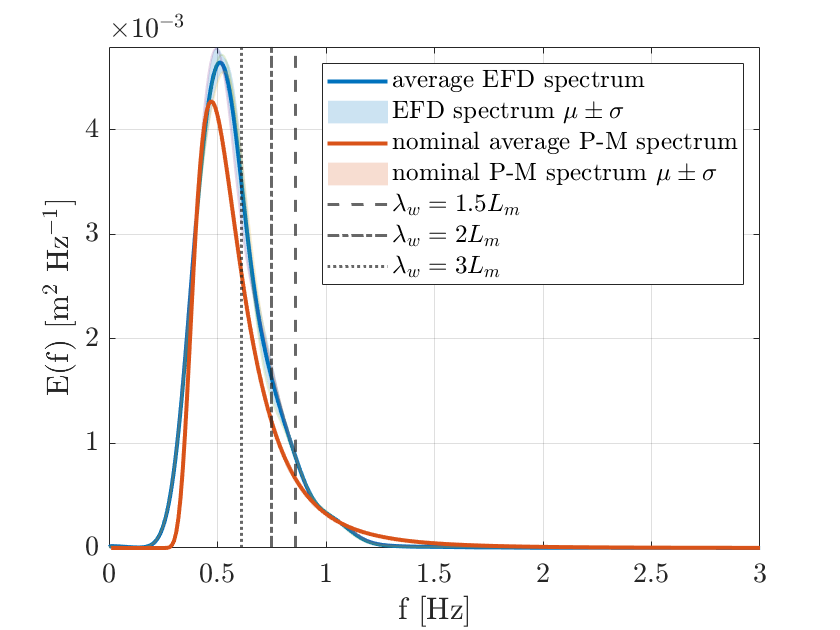}
    \caption{Comparison between nominal Pierson-Moskowitz and experimental power spectral density of wave elevation obtained from irregular wave test data. Vertical lines show the frequencies of regular waves tests.}
    \label{fig:pmspectrum}
\end{figure}
}

\subsubsection{Regular waves}
In the second part of the experiments, the moored model was subjected to head-sea regular wave trains. The nominal wave height $h$ was fixed to $h=0.1$~M model scale (5~m full scale) Three different wavelengths $\lambda_w = 1.5 L_{m}$, $2 L_{m}$, and $3L_{m}$ were tested. Actual realizations were characterized by (average $\pm$ standard deviation) $\hat h=0.105\pm0.01$~m, $0.107\pm0.01$~m, and $0.116\pm0.01$~m for the three $\lambda_w$, respectively.
%$\lambda_w = 1.52 L_{m}$, $2 L_{m}$, and $3.1L_{m}$ and 

\subsection{Instrumentation}

\subsubsection{Wave elevation measurements}
\begin{figure}
    \centering
    \captionsetup[subfigure]{justification=centering}    
    \begin{subfigure}[b]{0.4\linewidth} 
       \includegraphics[width=\linewidth]{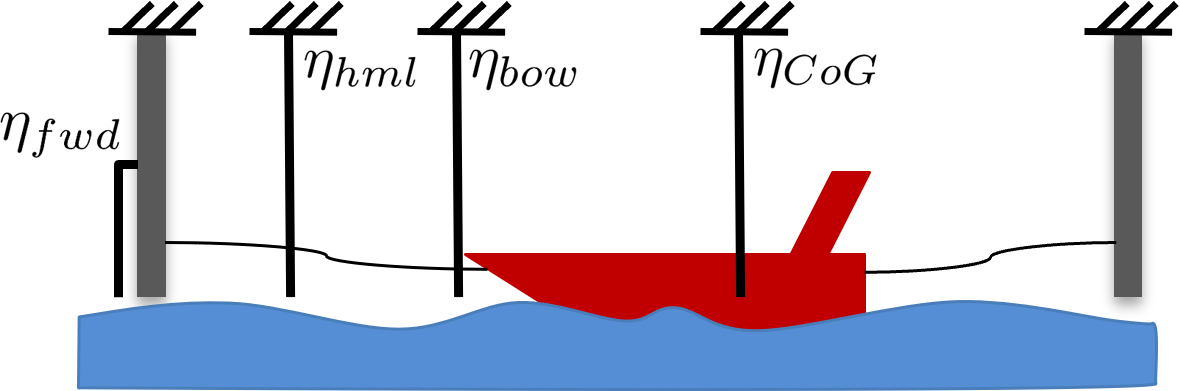}        \caption{}   \label{fig:mooredsketch}
    \end{subfigure}
    \hspace{1cm}
    \begin{subfigure}[b]{0.4\linewidth} 
    \begin{tabular}{ l l c }
        \midrule
        Type & Name & Position [m]\\
        \midrule
        Ultrasonic & $\eta^{us}_{fwd}$ & 48.42 \\
        Ultrasonic & $\eta^{us}_{hml}$ & 49.85 \\
        Ultrasonic & $\eta^{us}_\mathrm{bow}$ & 51.37 \\
        Ultrasonic & $\eta^{us}_{cog}$ & 52.09 \\
        Kenek & $\eta^{kn}_{fwd}$ & 48.42 \\
        Kenek & $\eta^{kn}_{cog}$ & 52.09 \\
        \midrule   
    \end{tabular}
    \caption{}\label{tab:waveprobes}
    \end{subfigure}
    \caption{Sketch of the wave probes location (\subref{fig:mooredsketch}), and wave probes location from wave generator flap (\subref{tab:waveprobes}).}\label{ft:waveprobes}  
\end{figure}

Two capacitance-type probes by Kenek (model SPH 150) were rigidly mounted to the carriage for measuring wave elevation at the forward mooring pole and the center of gravity of the model at rest. In addition, four ultrasonic wave elevation probes were also mounted, two of them in correspondence with the capacitance-type probes, one in correspondence with the half of the forward mooring line (at rest), and one in correspondence with the bow of the model at rest.
%accuracy, resolution, etc.
The distance in meters from the wave generator flap to each probe is reported in \cref{tab:waveprobes}.

The elevation probes collected data with a sampling frequency of 500~Hz, which were filtered by applying a low-pass minimum-order filter with delay compensation and stopband attenuation of 60 dB, with passband frequency $f_{pb}=10$~Hz and stopband frequency $f_{sb}=10.75$~Hz

\subsubsection{Motion measurements}
An SMC-108 motion reference unit (MRU) was installed onboard the CK-14e, composed of accelerometers and gyroscopes mounted in all three axes. 
The sensing data signals were acquired with a sampling frequency $f_s=50$~Hz.
Measures were processed online in parallel with a Kalman filter inside the MRU multi-core processor. 
This provided output data for accelerations $\ddot x, \ddot y, \ddot z$, angular velocities $\dot \phi, \dot \theta$, heave $z$, roll $\phi$, pitch $\theta$, and attitude $\psi$, with limited influence from accelerations and noise.
In particular, $z$ was internally computed by integration of $\ddot z$ into a linear position. 
The integration was further internally processed and filtered for an accurate heave measurement. 

The set of measured variables was further extended postprocessing the measurements, including $\dot x$ and $\dot z$, obtained by integrating the measured accelerations, and $\ddot \theta$, obtained by 4-th order central differentiation from the corresponding measured angular velocity. A low-pass minimum-order filter with a stopband attenuation of 60 dB and delay compensation was applied after the integration step, with passband frequency $f_{pb}=10$~Hz and stopband frequency $f_{sb}=10.75$~Hz.
%, along with detrending.

Additionally, a video recording of the moored setup from a side view was taken with a GoPro Hero 7 camera. 
A dedicated image-processing pipeline was developed to track the motion of the ship model within the towing tank from the video of each test case. Each video frame was first converted to the RGB color space and segmented using fixed intensity thresholds to isolate the region corresponding to a white identification panel positioned on top of the model to be used as the tracked object. The resulting binary mask was refined through morphological operations (opening, closing, and hole-filling) in order to remove noise and produce a clean and contiguous representation of the target. Connected-component analysis was then performed, and the panel was identified as the largest component. Its geometric properties — including centroid position and orientation angle — were extracted using standard region-based feature analysis. These measurements were stored frame by frame and subsequently low-pass filtered to suppress high-frequency noise and obtain estimations for the surge $x$, $z$, and $\theta$.
The pitch and heave from the image-processing tracking approach were compared with the MRU signals, obtaining a good agreement for the 2D motion variables, validating the procedure and also the surge measurement.
%$z$ and $\theta$. 

%The proposed approach is fully deterministic, computationally lightweight, and well-suited for laboratory environments with controlled illumination and a stationary background, making it a robust alternative to optical-flow or machine-learning-based tracking methods.
\subsubsection{Mooring forces}

Mooring forces $M_\mathrm{bow}$ and $M_\mathrm{stern}$ were measured by using two high-accuracy DDEN in-line submersible load cells by Applied Measurements Ltd., aligned with the mooring lines at calm water level. Pretensions have been applied accordingly to the elastic constants of the mooring lines and expected motions of the model in the test area.
%https://appmeas.co.uk/products/load-cells-force-sensors/in-line-submersible-load-cell-dden/

\section{HDMDc} \label{s:dmd}
DMD \cite{schmid2008dynamic,schmid2010} was originally presented to decompose high-dimensional time-resolved data into a set of spatiotemporal coherent structures, characterized by fixed spatial structures (modes) and associated temporal dynamics, providing a linear reduced-order representation of possibly nonlinear system dynamics.
The original DMD characterizes naturally evolving dynamical systems. In contrast, its extension to forced systems, called DMD with control (DMDc) \cite{proctor2016dynamic}, accounts for the influence of forcing inputs in the analysis, helping disambiguate it from the unforced dynamics of the system. 

The standard DMD and DMDc formulations approximate the Koopman operator, creating a best-fit linear model that links sequential data snapshots of measurements \cite{schmid2010,kutz2016dynamic,mezic2022}. This model provides a locally linear representation of the dynamics, which, however, is unable to capture many essential features of nonlinear systems.
The augmentation of the system state is thus the subject of several DMD algorithmic variants \cite{otto2019,takeishi2017,Lusch2018,Brunton2021} 
aiming to find a coordinate system (or \textit{embedding}) that spans a Koopman-invariant subspace, to search for an approximation of the Koopman operator valid also far from fixed points and periodic orbits in a larger space.
%The need for state augmentation through additional observables is even more critical for applications in which the number of states in the system is small, typically smaller than the number of available snapshots, such as the case at hand.
However, there is no general rule for defining these observables and guaranteeing they will form a closed subspace under the Koopman operator \cite{brunton2016b}.

The HDMD \cite{mezic2017} is a specific version of the DMD algorithm developed to deal with the cases of nonlinear systems in which only partial observations are available \cite{Brunton2021}.
Incorporating time-lagged information in the data used to learn the model, HDMD and HDMDc increase the dimensionality of the system. 
Including time-delayed data in the analysis, the HDMD and its extension to externally forced systems, HDMDc, can extract linear modes and the associated input operator defined in a space of augmented dimensionality. Such modes are capable of reflecting the non-linearities in the time evolution of the original system through complex relations between present and past states.
The state vector is thus augmented, embedding $s$ time-delayed copies of the original variables. 
The HDMDc involves, in addition, augmenting the input vector with $z$ time-delayed copies of the original forcing inputs. 
The use of time-delayed copies as additional observables in the DMD has been connected to the Koopman operator as a universal linearizing basis \cite{Brunton2017}, yielding the true Koopman eigenfunctions and eigenvalues in the limit of infinite-time observations \cite{mezic2017}.

%\subsubsection{HDMDc}\label{s:hdmdc}
%%%%%%%%%%%%%%%%%%%%%%%%%%%%
The HDMDc identifies a representation of the dynamics as an externally forced system:
\begin{equation}\label{eq:hdmdcdsys}
    \mathbf{\hat{x}}_{j+1} = \widehat{\mathbf{A}}\mathbf{\hat{x}}_j + \widehat{\mathbf{B}}\mathbf{\hat{u}}_j.
\end{equation}
The vectors $\mathbf{\hat{x}}_j$ and $\mathbf{\hat{u}}_j$ are called the extended state and input vectors snapshot at the time instant $j$.
The definition of $\mathbf{\hat{x}}_j$ and $\mathbf{\hat{u}}_j$ is obtained starting from the original state and input vectors snapshots $\mathbf{x}_j$ and $\mathbf{u}_j$, respectively:
\begin{alignat}{2}
    \mathbf{x}_j &= 
    \begin{bmatrix}
    x_{1} \\ x_{2} \\ \dots \\ x_{N}
    \end{bmatrix} \in \mathbb{R}^{N},
\quad
    \mathbf{u}_j &= 
    \begin{bmatrix}
    u_{1} \\ u_{2} \\ \dots \\ u_{Q}
    \end{bmatrix} \in \mathbb{R}^{Q},
\end{alignat}
which are augmented by embedding $s$ and $z$ time-lagged (delayed) copies of the original state and input variables, such that:
\begin{align}
    \hat{\mathbf{x}}_j = 
    \begin{bmatrix}
    \mathbf{x}_j \\ \mathbf{x}_{j-1} \\ \dots \\ \mathbf{x}_{j-s}
    \end{bmatrix} \in \mathbb{R}^{N(s+1)},
\quad
    \hat{\mathbf{u}}_j = 
    \begin{bmatrix}
    \mathbf{u}_j \\ \mathbf{u}_{j-1} \\ \dots \\ \mathbf{u}_{j-z}
    \end{bmatrix} \in \mathbb{R}^{Q(z+1)}.
\end{align}
As a consequence, the extended state matrix and the extended system input matrix are defined as
$\widehat{\mathbf{A}} \in \mathbb{R}^{N (s+1) \times N (s+1)}$ and $\widehat{\mathbf{B}} \in \mathbb{R}^{N (s+1) \times Q (z+1)}$, respectively. 

The procedure to extract the extended matrices from data starts by introducing the vector $\mathbf{\hat{y}}_j \in \mathbb{R}^{N(s+1)+Q(z+1)}$
\begin{equation}\label{eq:Y}
\mathbf{\hat{y}}_j=
\begin{bmatrix}
\mathbf{\hat{x}}_j \\
\mathbf{\hat{u}}_j\\
\end{bmatrix},
\end{equation}
such that \cref{eq:hdmdcdsys} can be rewritten as:
\begin{equation}\label{eq:dmdSIY}
\mathbf{\hat{x}}_{j+1}=\mathbf{\widehat{G}\hat{y}}_j, \hspace{1cm} \text{with} \hspace{1cm} \mathbf{\widehat{G}}=
\begin{bmatrix}
    \mathbf{\widehat{A}} & \mathbf{\widehat{B}} \\
\end{bmatrix}.
\end{equation}
Data are collected for model building from $m$ snapshots and are arranged in two augmented data matrices $\widehat{\mathbf{Y}} \in \mathbb{R}^{(N(s+1)+Q(z+1))\times (m-1)}$ and $\widehat{\mathbf{X}}' \in \mathbb{R}^{N(s+1)\times(m-1)}$, which are built as:
\begin{equation}\label{eq:scXX'}
\widehat{\mathbf{Y}}=
\begin{bmatrix}
\mathbf{X} \\
\mathbf{S}\\
\mathbf{U} \\
\mathbf{Z}\\
\end{bmatrix},
\qquad
\widehat{\mathbf{X}}'=
\begin{bmatrix}
\mathbf{X}' \\ 
\mathbf{S}'\\
\end{bmatrix}.
\end{equation}
The matrices $\mathbf{X} \in \mathbb{R}^{N \times (m-1)}$, $\mathbf{X}' \in \mathbb{R}^{N\times (m-1)}$, and $\mathbf{U} \in \mathbb{R}^{Q \times (m-1)}$ contain the extended state and input snapshots at the $m$ considered time instants:
\begin{equation}\label{eq:XX'}
\begin{split}
\mathbf{X}=
\begin{bmatrix}
\mathbf{x}_j & \mathbf{x}_{j+1} & \dots & \mathbf{x}_{m-1}\\
\end{bmatrix},\\
\mathbf{X}'=
\begin{bmatrix}
\mathbf{x}_{j+1} & \mathbf{x}_{j+2} & \dots & \mathbf{x}_{m}\\
\end{bmatrix},\\
\mathbf{U}=
\begin{bmatrix}
\mathbf{u}_j & \mathbf{u}_{j+1} & \dots & \mathbf{u}_{m-1}\\
\end{bmatrix},
\end{split}
\end{equation}
while the Hankel matrices $\mathbf{S}$, $\mathbf{S}'$, and $\mathbf{Z}$ contain the lagged (delayed) extended state and input snapshots:  
\begin{align}\label{eq:s}
\mathbf{S}&=
\begin{bmatrix}
\mathbf{x}_{j-1} & \mathbf{x}_{j} & \dots & \mathbf{x}_{m-2}\\
\mathbf{x}_{j-2} & \mathbf{x}_{j-1} & \dots & \mathbf{x}_{m-3}\\
\vdots & \vdots & \vdots & \vdots \\
\mathbf{x}_{j-s-1} & \mathbf{x}_{j-s} & \dots & \mathbf{x}_{m-s-1}
\end{bmatrix}, 
\end{align}
\begin{align}\label{eq:s'}
\mathbf{S}'&=
\begin{bmatrix}
\mathbf{x}_{j} & \mathbf{x}_{j+1} & \dots & \mathbf{x}_{m-1}\\
\mathbf{x}_{j-1} & \mathbf{x}_{j} & \dots & \mathbf{x}_{m-2}\\
\vdots & \vdots & \vdots & \vdots \\
\mathbf{x}_{j-s} & \mathbf{x}_{j-s+1} & \dots & \mathbf{x}_{m-s}
&\end{bmatrix},
\end{align}
\begin{align}\label{eq:z}
\mathbf{Z}&=
\begin{bmatrix}
\mathbf{u}_{j-1} & \mathbf{u}_{j} & \dots & \mathbf{u}_{m-2}\\
\mathbf{u}_{j-2} & \mathbf{u}_{j-1} & \dots & \mathbf{u}_{m-3}\\
\vdots & \vdots & \vdots & \vdots \\
\mathbf{u}_{j-z-1} & \mathbf{u}_{j-z} & \dots & \mathbf{u}_{m-z-1}\\
\end{bmatrix}.    
\end{align}
The augmented matrix $\widehat{\mathbf{G}} = [\widehat{\mathbf{A}}\; \widehat{\mathbf{B}}] \in \mathbb{R}^{N(s+1) \times(N(s+1)+Q(z+1))}$ is approximated by HDMDc by solving the following regularized least-square minimization:
\begin{equation}
    \min_{\widehat{A},\widehat{B}} \quad || \widehat{\mathbf{X}}' - \mathbf{[\widehat{\mathbf{A}}\; \widehat{\mathbf{B}}]} \; \widehat{\mathbf{Y}} ||^2_{F} + \lambda  \mathbf{[\widehat{\mathbf{A}}\; \widehat{\mathbf{B}}]} ||^2_{F},
\end{equation}
which solution is given by:
\begin{equation}
   [\widehat{\mathbf{A}}\; \widehat{\mathbf{B}}] = \widehat{\mathbf{X}}' \widehat{\mathbf{Y}}^{\text{T}} \left[ \widehat{\mathbf{Y}}\widehat{\mathbf{Y}}^{\text{T}} + \lambda \mathbf{I} \right]^{-1}.
\end{equation}
where $\lambda$ is a regularization factor.
The above Tikhonov-regularized formulation extends the one presented in \cite{XIE2024} to the HDMDc. It is suitable for improving the numerical stability of DMD regression, its robustness in high-dimensional spaces, and accuracy for applications involving noisy data. 
A similar effect can be pursued with the exact-DMD formulation by SVD rank truncation, see \citep{palmaFAST2025}, also reducing the model sizes. The Tikhonov-regularized formulation is here preferred for its lower computational cost, as it does not involve SVD of the snapshots matrices, and for its more robust behavior to noisy data.
%è una tecnica ben consolidata per affrontare problemi ill-posed e prevenire l'overfitting nei modelli di regressione. La sua applicazione nella DMDc è una naturale estensione per migliorare la robustezza del modello.
Once the matrices $\widehat{\mathbf{A}}$ and $\widehat{\mathbf{B}}$ are obtained, \cref{eq:hdmdcdsys} can be used to calculate the time evolution of the augmented state vector from an initial condition $\mathbf{\tilde{\hat x}}(t)$, where the tilde indicates the HDMDc estimation. By isolating its first $N$ components, the predicted time evolution of the original state variables $\mathbf{\tilde{x}}(t)$ is extracted.

\subsection{Uncertainty quantification in HDMDc through ensembling}
DMD-based models can be further extended to provide some uncertainty estimation of their predictions.
To this aim, the ensembling approach is applied, \textit{i.e.,} the combination of predictions coming from different models to obtain a prediction along with its statistics.
The Bayesian formulation is used here, first introduced in \cite{palma2025windturbine}, also adopted in \cite{palma2024forecasting} and \cite{palma2025si} for both HDMD and HDMDc, which leads to BHDMD and BHDMDc, respectively.
The driving rationale behind the BHDMDc is the observation highlighted by the authors in previous works \cite{serani2023, Diez2024, Serani2024snh} that the final prediction from DMD-based models can strongly vary for different hyperparameter settings of the method. 
Furthermore, there is no general rule for determining their optimal values. 

The dimensions and the values within matrices $\widehat{\mathbf{A}}$ and $\widehat{\mathbf{B}}$ depend on four hyperparameter, namely $m$, $s$, $z$ and $\lambda$. The first three can be conveniently expressed as time lengths, \textit{i.e.,}  the observation time length $l_{tr} = t_m - t_1$, the maximum delay time in the augmented state $l_{d_x} = t_{j-1} - t_{j-s-1}$, and the maximum delay time in the augmented input $l_{d_u} = t_{j-1} - t_{j-z-1}$.
These dependencies can be denoted as follows:
\begin{equation}\label{eq:bayes1}
\begin{split}
\quad \widehat{\mathbf{A}}&=\widehat{\mathbf{A}}(l_{tr},l_{d_x},l_{d_u},\lambda), \qquad \\
\widehat{\mathbf{B}}&=\widehat{\mathbf{B}}(l_{tr},l_{d_x},l_{d_u},\lambda).
\end{split}
\end{equation}
In BHDMDc, the hyperparameters are considered as stochastic variables with given prior distribution, $p(l_{tr})$, $p(l_{d_x})$, $p(l_{d_u})$, and $p(\lambda)$, respectively.
Through uncertainty propagation, the solution $\mathbf{\tilde x}(t)$ also depends on $l_{tr}$, $l_{d_x}$, $l_{d_u}$, and $\lambda$:
\begin{equation}\label{eq:bayes2}
\mathbf{x}=\mathbf{x}(t,l_{tr},l_{d_x},l_{d_u},\lambda),
\end{equation}
and the following are used to define, at a given time $t$, the expected value ($\mu$) of the solution and its standard deviation ($\sigma$):
\begin{equation}\label{eq:bayes3}
\mu[\mathbf x(t)]=\int_{\lambda^l}^{\lambda^u}\int_{l_{d_u}^l}^{l_{d_u}^u} \int_{l_{d_x}^l}^{l_{d_x}^u} \int_{l_{tr}^l}^{l_{tr}^u}\mathbf{x}(t,l_{tr},l_{d_x},l_{d_u},\lambda)p(l_{tr})p(l_{d_x})p(l_{d_u})p(\lambda)dl_{tr} dl_{d_x} dl_{d_u} d\lambda,
\end{equation}
\begin{equation}\label{eq:bayes4}
\sigma[\mathbf x(t)]= \left\{ \int_{\lambda^l}^{\lambda^u}\int_{l_{d_u}^l}^{l_{d_u}^u} \int_{l_{d_x}^l}^{l_{d_x}^u} 
 \int_{l_{tr}^l}^{l_{tr}^u} \left\{\mathbf{x}\left(t,l_{tr},l_{d_x},l_{d_u},\lambda\right)-\mu\left[\mathbf x\left(t\right)\right] \right\}^2 p(l_{tr})p(l_{d_x})p(l_{d_u})p(\lambda)dl_{tr} dl_{d_x} dl_{d_u} d\lambda \right\}^\frac{1}{2},
\end{equation}
where ${l_{tr}^l}$, ${l_{d_x}^l}$, ${l_{d_u}^l}, \lambda^l$ and ${l_{tr}^u}$, ${l_{d_x}^u}$, ${l_{d_u}^u}$, $\lambda^u$ are lower and upper bounds and $p(l_{tr})$, $p(l_{d_x})$, $p(l_{d_u})$ $p(\lambda)$, are the given probability density functions for $l_{tr}$, $l_{d_x}$, $l_{d_u}$, and $\lambda$.

In practice, a uniform probability density function is assigned to the hyperparameters, and a set of realizations is obtained through a Monte Carlo sampling, obtaining a posterior distribution on the prediction.

We note that the proposed method is termed "Bayesian" in a broader sense than classical Bayesian inference. In particular, posterior distributions are not inferred over the Koopman operator or dynamic mode decomposition matrices themselves. 
Instead, in BHDMDc, the hyperparameters of the method are treated in a Bayesian manner, assigning them given prior distributions and propagating their uncertainty through Monte Carlo sampling. This results in a predictive distribution for the system variables. 
While this does not constitute full Bayesian parameter inference, it is consistent with Bayesian principles of uncertainty propagation and model averaging, and aligns with how “Bayesian” is used in related literature for uncertainty-aware reduced-order modeling.
% It is worth noting that Bayesian methodologies in the strict sense refer to posterior inference on model parameters, typically involving Bayes’ theorem with explicit prior distributions and data likelihoods. However, in the present work, the term Bayesian is used in a broader but still methodologically grounded sense. Specifically, we adopt a Bayesian treatment of hyperparameters, namely the training window length and time delays in the Hankel-DMDc formulation, by modeling them as random variables with uniform priors, and performing Monte Carlo uncertainty propagation through the model. While this does not constitute full Bayesian inference (e.g., no posterior over the operator matrices $\bf A$, $\bf B$), it does follow Bayesian statistical principles by integrating over uncertainty in model inputs to produce a predictive distribution.

\section{Statistical variables and performance metrics}
%Statistical quantities (EV, std, quantiles, uncertainty, pdf and kernel density estimation)
% Statistical variables of interest are the expected value ($EV$), standard deviation ($SD$), probability 
% density function ($PDF$), and quantiles ($q$). For a given variable evolving in time, they are evaluated numerically using a 
% sample of $n$ items, extracted from the 
% time series as $J_i=J(t_i)$, $i=1,\dots,n$.

To evaluate the predictions made by the deterministic and Bayesian models and compare them with the ground truth from the experiments, three error indices are employed: the average normalized mean square error (ANRMSE) \citep{Diez2024}, its time-resolved version ($\varepsilon$), and the normalized average minimum/maximum absolute error (NAMMAE) \citep{Diez2024}. 

The ANRMSE quantifies the root mean square error between the predicted values $\mathrm{\mathbf{\tilde x}}_t$ and the measured (reference) values $\mathrm{\mathbf{x}}_t$ at different time steps, normalizing the result for each variable by $k$ times the standard deviation of the measured value ($k=1$ in this work), and averaging over the N variables in $\mathbf{x}$:
\begin{equation}\label{eq:nrmse}
   \mathrm{ANRMSE} = \frac{1}{N} \sum_{i=1}^{N} \sqrt{\frac{1}{\mathcal{T} (k\,SD[x_i])^2} \sum_{j=1}^{\mathcal{T}} \left( \tilde{x}_{i}(t_j) - x_{i}(t_j) \right)^2},
\end{equation}
where $\mathcal{T}$ is the number of considered time instants in the considered time window and $SD[x_i]$ is the standard deviation of the measured values in the considered time window for the variable $x_i$:
\begin{equation}
    SD[\xi] = \sqrt{\frac{1}{\mathcal{T}-1} \sum_{i=1}^\mathcal{T}(\xi_i-EV[\xi])^2}
\end{equation}
and
\begin{equation}
    EV[\xi] = \frac{1}{\mathcal{T}}\sum_{i=1}^\mathcal{T} \xi_i
\end{equation}
For the same time window, the time-resolved variant of the ANRMSE called $\varepsilon$ is evaluated as the average on the variables of the time evolution of the square root difference between the reference and predicted signal, normalized by its standard deviation in the considered time interval:
\begin{equation}
    \varepsilon(t) = \frac{1}{N} \sum_{i=1}^N \sqrt{\frac{\left(\tilde{x_i}(t)-x_i(t)\right)^2}{(k \, SD[x_i])^2}}
\end{equation}
%sum(sqrt(((xt1-Xte1).^2)./((knorm)^2*var(Xte1,1,2))),1)/nvar
This is used to monitor potential trends in the prediction error, \textit{i.e.,} whether the accuracy decreases for longer predictions.

The NAMMAE metric, introduced in \citet{diez2022snh,Diez2024}, provides an engineering-oriented assessment of prediction accuracy. It measures the absolute difference between the minimum and maximum values of the predicted and measured time series, normalized by $k$ times the standard deviation of the measured values ($k=1$ in this work), and averaged over $N$ variables, as follows:
\begin{equation}\label{eq:nammae}
    \mathrm{NAMMAE} = \frac{1}{2 N } \sum_{i=1}^{N} \frac{1}{k\,SD[x_i]}\Bigg( \left| \min(\tilde{x}_{i}(t_j)) - \min(x_{i}(t_j))\right| 
    + \left| \max(\tilde{x}_{i}(t_j)) - \max(x_{i}(t_j)) \right| \Bigg) \quad j=1,\dots,\mathcal{T}.
\end{equation}

In addition to the direct comparison of DMD-predicted and experimentally measured time histories, calculating the probability density function (PDF) of the variables under prediction is of interest for the application in irregular waves.
To statistically assess the PDF estimator obtaining confidence intervals, a moving block bootstrap (MBB) method is applied to time histories from EFD and BHDMDc-based predictions. 
%of each state variable for the EFD and the mean prediction of FHDMDc.
For the MBB analysis, a single long time history of $\mathcal{T}$ samples is obtained for the EFD and the BHDMDc, respectively, by joining all the measured and predicted test time series.
Starting from this time signal with $\mathcal{T}$ samples, 
a number $C = \mathcal{T}-r+1$ of moving blocks is used, each defining a time history with $\xi_i$ time samples $i=c,\dots,c+r-1$, where $c$ is the block index and $r = (2\varphi/a)^{2/3}\mathcal{T}^{1/3}$ is an optimal block length \cite{carlstein1986use} with
\begin{equation}
    \varphi = \dfrac{\mathcal{T}\displaystyle\sum_{i=1}^{\mathcal{T}} \left[ \xi_{i+1} - \text{EV}(\xi) \right] \left[ \xi_{i} - \text{EV}(\xi) \right] }{(\mathcal{T}-1)\displaystyle\sum_{i=1}^{\mathcal{T}} \left[\left( \xi_{i} - \text{EV}(\xi) \right) \right]}
\end{equation}
and $a=(1-\varphi)(1+\varphi)$. 
From the original set of C blocks, a number of $C' = \mathcal{T}/r$ blocks are drawn at random with replacement and concatenated in the order they are picked, forming a new bootstrapped series of size $\mathcal{T}$. 
The PDF of each bootstrapped time series %for both the EFD and mean FHDMDc predictions 
is obtained using kernel density estimation \cite{Miecznikowski2010} as follows:
\begin{equation}
    \text{PDF}\left(\xi,y\right) = \frac{1}{\mathcal{T} h}\sum_{i=1}^{\mathcal{T}} K \left(\frac{y -\xi_i}{h}\right).
\end{equation}
Here, $K$ is a normal kernel function defined as
\begin{equation}
    K\left(\xi\right) = \frac{1}{\sqrt{2 \pi}} \exp{\left(-\frac{\xi^2}{2}\right)},
\end{equation}
where 
\begin{equation}
h=1.06 \min\left(SD[\xi],\text{IQR}(\xi)\right)\mathcal{T}^{-1/5}
\end{equation}
is the bandwidth \cite{Silverman2018}, and $\text{IQR}(\xi)$ is the inter-quartile range for the variable $\xi$
\begin{equation}
    IQR[\xi] = q_\xi(0.75)-q_\xi(0.25)
\end{equation}
being $q_{\xi}$ the quantile function fpor the variable $\xi$.
%A total of B=100 bootstrapped series are used here, and hence, a set of 100 PDFs is obtained for each variable of the system state $\mathbf{x}$. 
The expected value and a confidence interval are calculated for the PDF of each variable for the EFD measurements and DMD-based predictions.
The quantile function $q$ is evaluated at probabilities $p=0.025$ and $0.975$, defining the lower and upper bounds of the 95\% confidence interval of the PDFs as $U_{\text{PDF}(\xi,y)} = q_{\text{PDF}(\xi,y)}(0.975) - q_{\text{PDF}(\xi,y)}(0.025)$. 

The so-obtained PDFs of each variable from the different sources are then compared using the Jensen-Shannon divergence (JSD) \cite{marlantes2024}, defined as:
\begin{align}
    &\mathrm{JSD}_{\xi} = \mathrm{JSD}[V_\xi||W_\xi] = \frac{1}{2}D(V_\xi\,||\,M_\xi) + \frac{1}{2}D(W_\xi\,||\,M_\xi),   \label{eq:jsd}
\end{align}
where
\begin{equation}
    M_\xi = \frac{1}{2} (V_\xi + W_\xi)\label{eq:jsd2}, \quad \text{and} \quad D(V_\xi\,||\,W_\xi)=\sum_{y \in \chi} V_\xi(y) \ln\left( \frac{V_\xi(y)}{W_\xi(y)} \right).
\end{equation}
The Jensen–Shannon divergence (JSD) \cite{Lin1991} is a symmetric measure of similarity between two probability distributions $V$ and $W$, based on the Kullback–Leibler divergence ($D$) \cite{Kullback1951}. 
Here, $V_\xi$ and $W_\xi$ are the PDFs of the variable $\xi$ from the EFD and HDMDc, respectively.
The JSD quantifies the average discrepancy of each distribution with respect to their mixture $M$, defined over the domain $\chi$. It is always finite, bounded as $0\le \mathrm{JSD}[V||M]\le \ln{2}$.    

For each variable, the expected value and the quantile function for $p=0.025$ and $0.975$ of JSD are calculated on the PDFs evaluated from the bootstrapped time series,
defining the lower and upper bound of the 95\% confidence interval $U_{\text{JSD}_{\xi}} =U_{\text{JSD}(V||W)} = \text{JSD}(V||W)_{q=0.95} - \text{JSD}(V||W)_{q=0.025}$.

\section{System identification setup}
% The HDMDc is used as a data-driven system identification method to obtain a reduced-order model for the CK-14e from experimental measurements.
{\color{black}In the present work, no equation-based motion model of the ASV is derived or assumed. The reduced-order model is obtained in a fully data-driven and equation-free manner, where the state-space representation of \cref{eq:hdmdcdsys} is identified directly from the experimental measurements through the (Bayesian) Hankel-DMDc (see Box~\ref{alg:det-hdmd-si} for a summary of the workflow for the system identification using HDMDc).
The state vector of the system is defined as
\begin{equation}
    \mathbf{x} = \left[ x, z ,\theta, \dot x, \dot z, \dot \theta, \ddot x, \ddot z, \ddot \theta, M_\mathrm{bow}, M_\mathrm{stern} \right]^T.
\end{equation}
The variables in the state vector correspond to the surge, heave, and pitch degrees of freedom, which dominate the symmetric head-sea response of the moored vessel, together with their first and second time derivatives and the measured mooring loads.
These quantities reflect the relevant physics of the phenomenon to be predicted by the system identification procedure.

The input vector is composed of observables based on the wave elevations measured by the Kenek probes $\eta^{kn}_{fwd}$ and $\eta^{kn}_{cog}$, and the model surge $x$. 
\begin{equation}
    \mathbf{u} = \left[ \eta_{fwd}, \eta_{cog} \right].
\end{equation}
 }
During the experiments, the relative position between the model and the wave probes was not constant, and large deviations from the rest position occurred in the x-direction.
The elastic restoring force from the moorings was not sufficient to counteract the thrust by the waves and keep the surge oscillating around the rest position.
Hence, directly using the wave elevation in the input vector as measured by the probes would lead to non-negligible phase error in the prediction. 
The HDMDc and BHDMDc are, in fact, able to learn only a constant phase relation between input and output, which, however, would be insufficient due to the dynamic change of the relative position between probes and model.
For this reason, the observables $\eta_{fwd}$ and $\eta_{cog}$ were obtained as delayed signals from $\eta^{kn}_{fwd}$ and $\eta^{kn}_{cog}$: 
\begin{equation}
\begin{split}
    \eta_{fwd}(t) &= \eta_{fwd}^{kn}(t-\tau(t))\\
    \eta_{cog}(t) &= \eta_{cog}^{kn}(t-\tau(t))
\end{split}
\end{equation}
where the delay $\tau$ depends on the surge $x$ and the wave celerity $c$:
\begin{equation}
    \tau(t) = \frac{x(t)-x_0}{c}.
\end{equation}
In this work, a simple estimation of $c$ is obtained through discrete signal correlation between $\eta_{fwd}^{kn}$ and $\eta_{cog}^{kn}$ as \textit{effective} phase velocity:
\begin{equation}
    c = \frac{x^{kn}_{cog}-x^{kn}_{fwd}}{\Delta t_s}, \quad \Delta t_s = \operatorname*{argmax}_{\Delta t_s}\left\{\text{corr}[\eta_{fwd}(t), \eta_{cog}(t-\Delta t_s)]\right\}
\end{equation}
{\color{black}The discrete delay $\Delta t_s$ is defined as the temporal shift that maximizes the normalized cross-correlation between the two signals.}

An average encounter wave period $\hat{T}=1.64s$ is calculated as the average encounter wave period of the measured $\eta^{kn}_{\text{fwd}}$ signal in the three EFD runs with irregular waves:
\begin{equation}
    \hat{T} = \frac{1}{n_\text{EFD}}\sum_{i=1}^{n_\text{EFD}}\frac{N{z_{up_i}}}{\mathcal{T}_i}.
\end{equation}

For processing ease with DMD, data were downsampled to 32 time steps per $\hat{T}$. This reduced the size of the data matrices to be handled while preserving sufficient temporal resolution to retain the fidelity of the original signal.

Exploiting results from previous studies by the authors \cite{palma2025si, palmaFAST2025,palma2025sicatamaran}, reasonable values for the HDMDc hyperparameters for deterministic analysis are derived: $l^{det}_{tr} = 20 \hat{T}$, $l^{det}_{d_x} = \hat{T}$, $l^{det}_{d_u} = 5\hat{T}$ and $\lambda^{det} = 100$, so that $m^{det}=640$, $s^{det}=32$, and $z^{det}=160$.

The ranges of variation for the uniform hyperparameter distributions in the Bayesian analysis were obtained considering a $\pm 50\%$ interval centered on the deterministic values, defining the respective priors:  
$l_{tr}^{bay}$~$\sim$~$\mathcal{U}(10\hat T,30 \hat T)$,
$l_{d_x}^{bay}$~$\sim$~$\mathcal{U}(0.5\hat T, 1.51\hat T)$,
$l_{d_u}^{bay}$~$\sim$~$\mathcal{U}(2.5\hat T,7.5 \hat T)$,
and $\lambda^{bay}$~$\sim$~$\mathcal{U}(50,150)$.
The Bayesian predictions are obtained using 100 Monte Carlo realizations of the hyperparameters.
The actual number of delays $s$ and $z$ and training samples $m$ are taken as the nearest integers from the calculated values.

All the analyses are based on normalized data using the Z-score standardization. Specifically, the time histories of each variable are shifted and scaled using the average and standard deviation evaluated on the training signal.

\begin{footnotesize}
\begin{algobox}[label=alg:det-hdmd-si]{System identification using HDMDc workflow}
\textbf{Inputs.} Training state and input time series $\mathbf{x}(t)\in\mathbb{R}^N$, $\mathbf{u}(t)\in\mathbb{R}^Q$; 
training window length $l_{tr}$; delay-embedding length $l_d$; 
Testing input time series $\mathbf{u}(t)\in\mathbb{R}^Q$ for the forecast horizon $l_{te}$.

\begin{enumerate}[noitemsep]
    \item \textbf{Preprocessing.} 
    \begin{itemize}[noitemsep]
        \item Identify the training window of length $l_{tr}$ within the reference time series; 
        \item (optional) subsample time series;
        \item apply z-scoring to $\mathbf{x}(t)$ and $\mathbf{u}(t)$.
    \end{itemize}    

    \item \textbf{Build snapshots matrices.} 
    \begin{itemize}[noitemsep]
        \item With $m=\lfloor l_{tr}/\Delta t \rfloor$, form $\mathbf{X}=[\mathbf{x}_j,\dots,\mathbf{x}_{m-1}]$, $\mathbf{X}'=[\mathbf{x}_{j+1},\dots,\mathbf{x}_m]$, and $\mathbf{U}=[\mathbf{u}_j,\dots,\mathbf{u}_{m-1}]$ from the training data.
    \end{itemize}

    \item \textbf{Hankel embedding.} 
    \begin{itemize}[noitemsep]
        \item With $s=\lfloor l_{dx}/\Delta t \rfloor$ and $z=\lfloor l_{du}/\Delta t \rfloor$, build delayed state and input vectors:
        \[
        \mathbf{\widehat{x}}_j = [\mathbf{x}_j^T, \mathbf{x}_{j-1}^T, \dots, \mathbf{x}_{j-s}^T]^T, \quad 
        \mathbf{\widehat{u}}_j = [\mathbf{u}_j^T, \mathbf{u}_{j-1}^T, \dots, \mathbf{u}_{j-z}^T]^T;
        \]
        \item assemble extended snapshot matrices $\mathbf{\widehat{X}}, \mathbf{\widehat{X}}'$, and $\mathbf{\widehat{U}}$.
    \end{itemize}

    \item \textbf{HDMDc operator estimation.} 
    \begin{itemize}[noitemsep]
        \item Define $\mathbf{\widehat{Y}} = \begin{bmatrix}\mathbf{\widehat{X}} \\ \mathbf{\widehat{U}}\end{bmatrix}$;
        \item estimate the augmented operator by solving the Tikhonov-regularized least-square problem:
        \[
        [\mathbf{\widehat{A}}\;\mathbf{\widehat{B}}] = \mathbf{\widehat{X}}' \mathbf{\widehat{Y}}^T 
        \left(\mathbf{\widehat{Y}}\mathbf{\widehat{Y}}^T + \lambda \mathbf{I}\right)^{-1}.
        \]
    \end{itemize}

    \item \textbf{Forecasting.} 
    \begin{itemize}[noitemsep]
        \item Collect $\hat{u}(t)$ for the test window $l_{te}$;
        \item for each time step $k \in [0, l_{te}]$, predict $\mathbf{\tilde{\hat{x}}}$
        \[
        \mathbf{\tilde{\hat{x}}}_{k+1} = \mathbf{\widehat{A}}\mathbf{\tilde{\hat{x}}}_k + \mathbf{\widehat{B}}\mathbf{\hat{u}}_k;
        \]
        \item extract original variables $\mathbf{\tilde{x}}(t)$ from the first $N$ components of $\mathbf{\tilde{\widehat{x}}}(t)$;
        \item (optional) de-standardize using $\mu_{tr}, \sigma_{tr}$.
    \end{itemize}
\end{enumerate}

\textbf{Outputs.} Deterministic prediction $\mathbf{\tilde{x}}(t)$ for the horizon $l_{te}$ and identified system matrices $\mathbf{\widehat{A}}, \mathbf{\widehat{B}}$.
\end{algobox}
\end{footnotesize}

This paper aims to assess the capability of HDMDc and BHDMDc as data-driven system identification methods for obtaining a reduced-order model of the CK-14e from experimental measurements, predicting the vessel response to waves not only under the same conditions used for learning the model, but also in unseen conditions. 
For this reason, the data from each experimental run in irregular waves was split into \textit{training} (the first half) and \textit{testing} data (the second half). Data from regular wave tests are used completely as \textit{testing}.

%%%%%%%%%%%%%%%%%%%%%%%%%%%%%%%%%%%%%%%%%%
\section{Results and discussion}\label{s:res}
The main scope of the article is to investigate the capability of HDMDc and BHDMDc-based ROMs to accurately predict the ship's response to unseen wave inputs, also in sea conditions that differ from the ones used in model learning.
Hence, data from sea state 7 P-M spectrum head waves are used for training, and then the so obtained ROMs are first applied to unseen input waves following the same P-M sea spectrum, and then to regular wave conditions.
Results are consequently structured in two parts, namely irregular-to-irregular and irregular-to-regular, reflecting the two testing conditions.

In order to statistically assess the performance of HDMDc and BHDMDc, the ANRMSE and NAMMAE were evaluated for several training and testing sequences. In particular, 10 training sequences and 10 test sequences randomly selected were combined in a full-factorial manner. The same training and test sequences were used for the deterministic and Bayesian versions of the ROM to guarantee a fair comparison.
The length of the test time histories for the irregular waves case was $l_{te}=15\hat{T}$. Due to reduced signal lengths, $l_{te}=5\hat{T}$ for regular waves.

For both irregular-to-irregular and irregular-to-regular cases, results are presented as box-violin plots comparing the ANRMSE and NAMMAE for the deterministic and Bayesian ROMs, see \cref{fig:irr2irr_boxviolin} and \cref{fig:irr2reg_boxviolin}, respectively.
The boxes show the first, second (equivalent to the median value), and third quartiles, while the whiskers extend from the box to the farthest data point lying within 1.5 times the interquartile range, defined as the difference between the third and the first quartiles from the box. 
The density of the data distribution is additionally represented by the violin shape, providing a visual summary of the distribution beyond the quartiles.
The values of the single realizations, including outliers, are plotted with colored dots.

In addition, representative test sequences are shown in \cref{fig:irr2irr1,fig:irr2irr2} for irregular-to-irregular and in \cref{fig:irr2reg1,fig:irr2reg2,fig:irr2reg3} for irregular-to-regular cases. Figures compare the experimental measurements (EFD), the deterministic ROM prediction, and the Bayesian ROM prediction (mean value as solid line and 95\% confidence interval as shaded area of the same color).

\begin{figure}[ht!]
    \centering
    \includegraphics[width=0.75\linewidth]{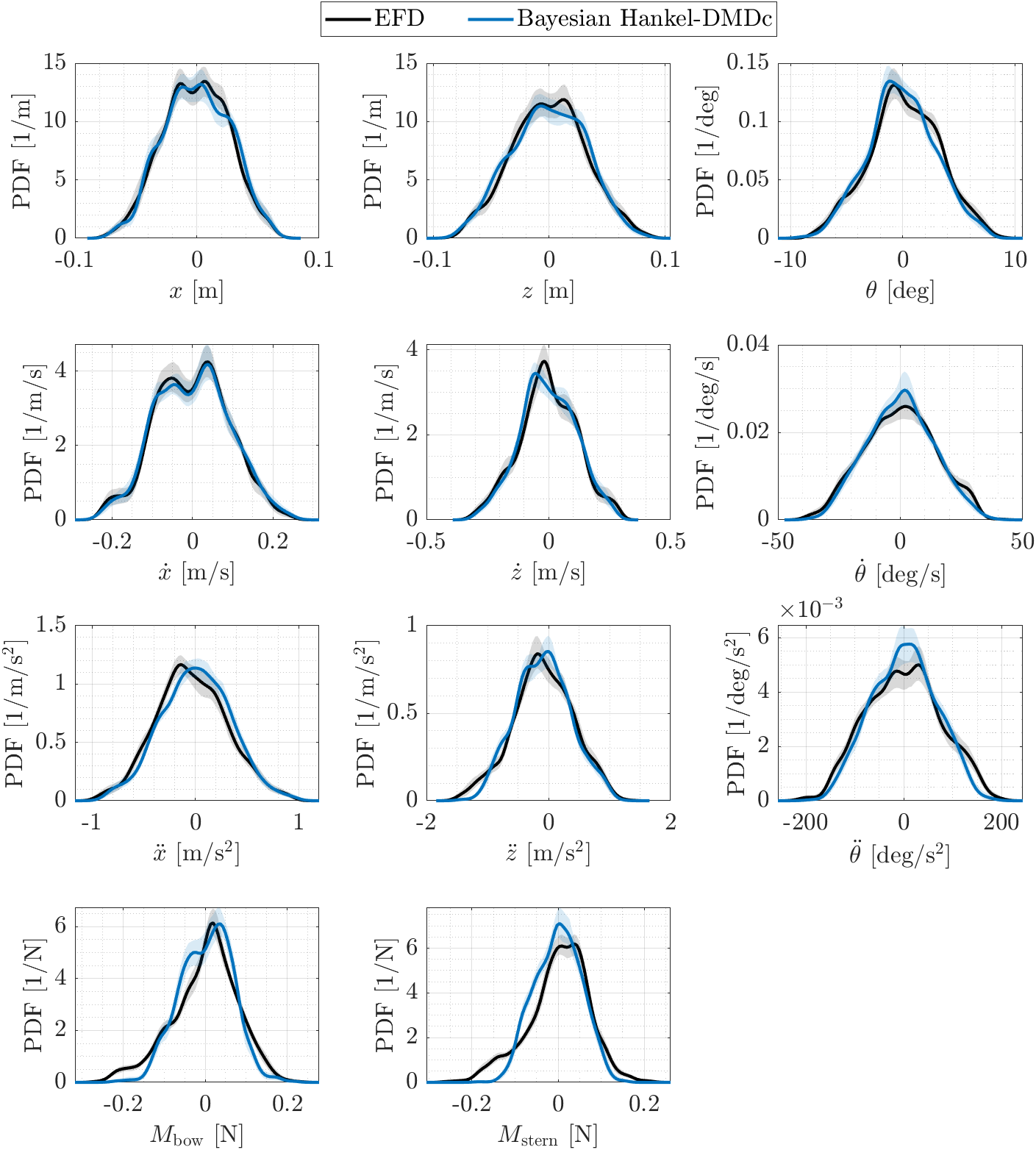}
    \caption{PDF comparison between measured data and the EV of the Bayesian Hankel-DMDc prediction on bootstrapped sequences for each variable. Shaded areas indicate the 95\% confidence interval of the two PDFs.}
    \label{fig:pdfbootstrap}
\end{figure}

Finally, the MBB analysis is applied to the irregular-to-irregular case, and PDFs from bootstrapped sequences for EFD and BHDMDc are obtained for each predicted variable, along with their 95\% confidence interval, and shown in \cref{fig:pdfbootstrap}.
The differences between the EFD and BHDMDc distributions are evaluated using JSD, and results are summarized in \cref{tab:jsd_bootstrap}.

\subsection{Irregular-to-irregular}
The HDMDc and the BHDMDc ROMs produce an overall reliable and robust prediction for unseen irregular input waves.
No trend in $\varepsilon(t)$ was observed throughout the testing window, suggesting that the same level of accuracy shown in $t\le15\hat T$ can be achieved for arbitrarily long sequences.

\begin{figure}[ht!]
    \centering
    \captionsetup[subfigure]{justification=centering}    
    \begin{subfigure}[b]{0.49\linewidth} 
        \includegraphics[width=\linewidth]{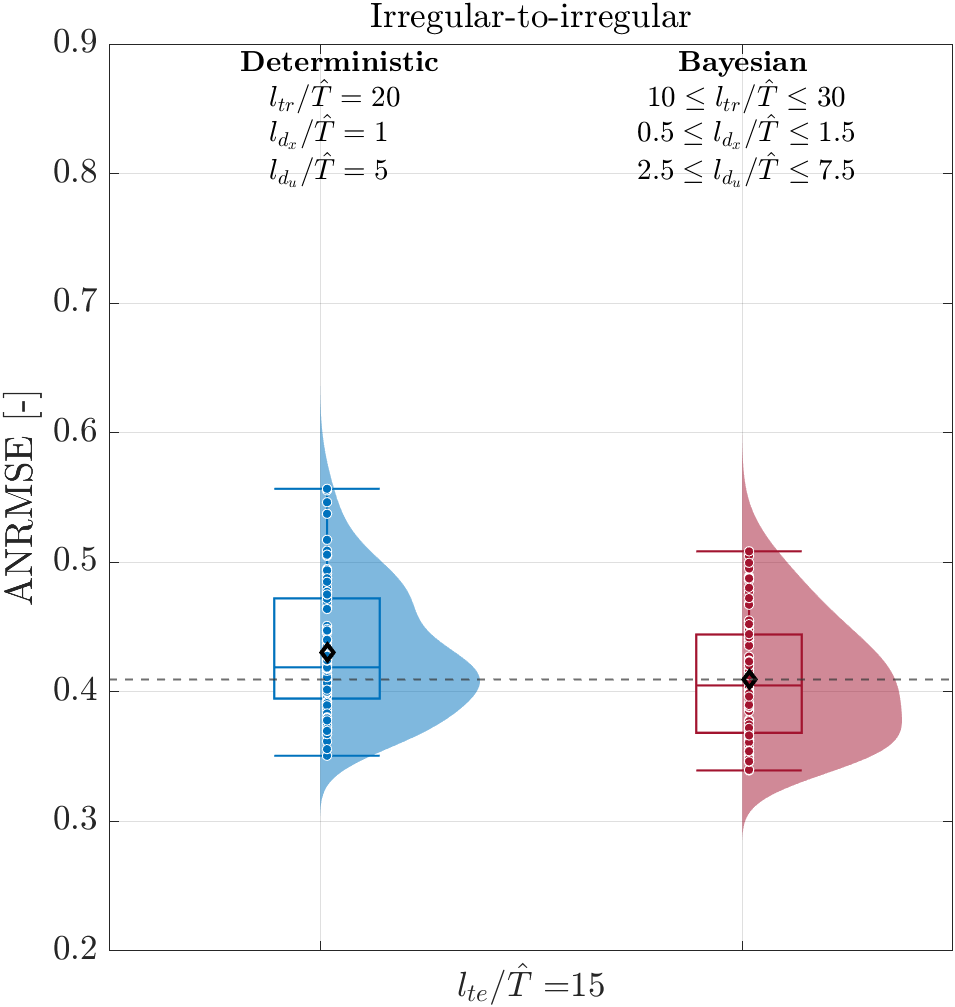}
        \caption{} \label{fig:irr2irrANRMSE}
    \end{subfigure}
    \hfill
    \begin{subfigure}[b]{0.49\linewidth} 
        \includegraphics[width=\linewidth]{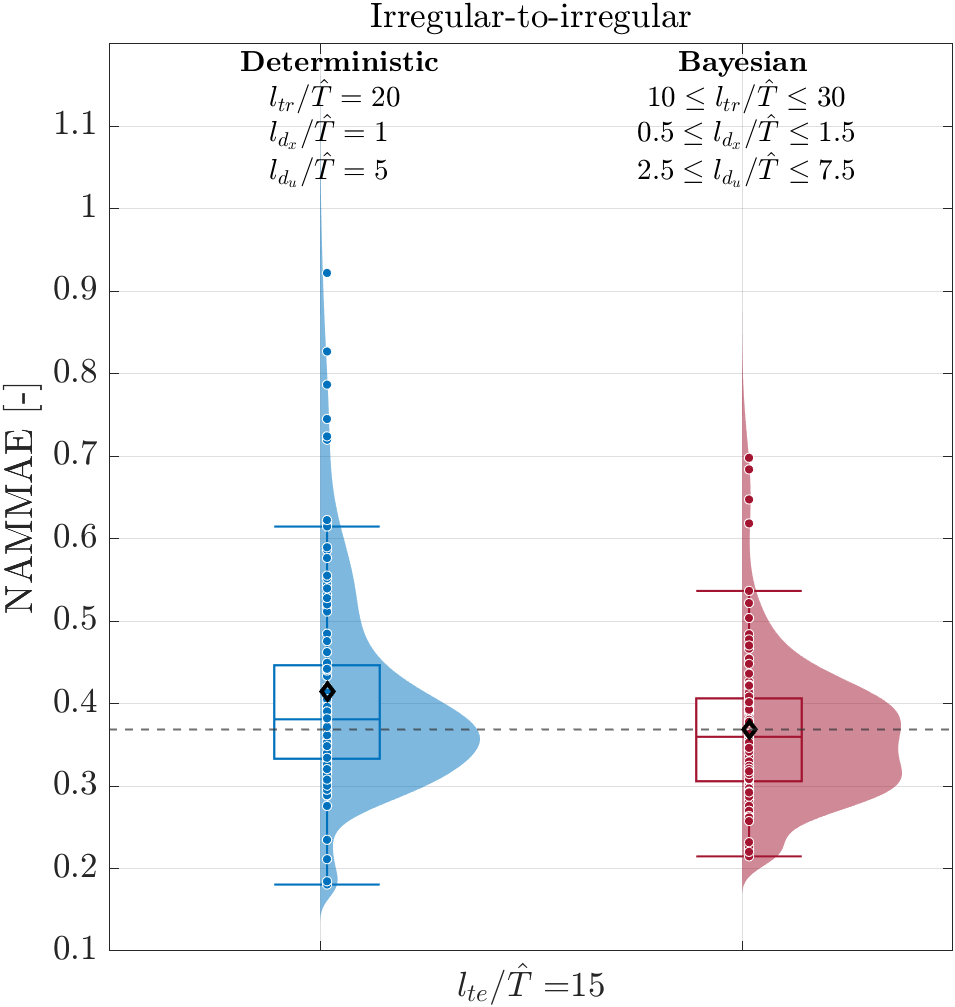}
        \caption{}\label{fig:irr2irrNAMMAE}
    \end{subfigure}
    \caption{ANRMSE (\subref{fig:irr2irrANRMSE}) and NAMMAE (\subref{fig:irr2irrNAMMAE}) box-violin plots. Comparison between deterministic and Bayesian predictions over the test sequences.}
    \label{fig:irr2irr_boxviolin}
\end{figure}

\begin{figure}[ht!]
    \centering
    \includegraphics[width=0.9\linewidth]{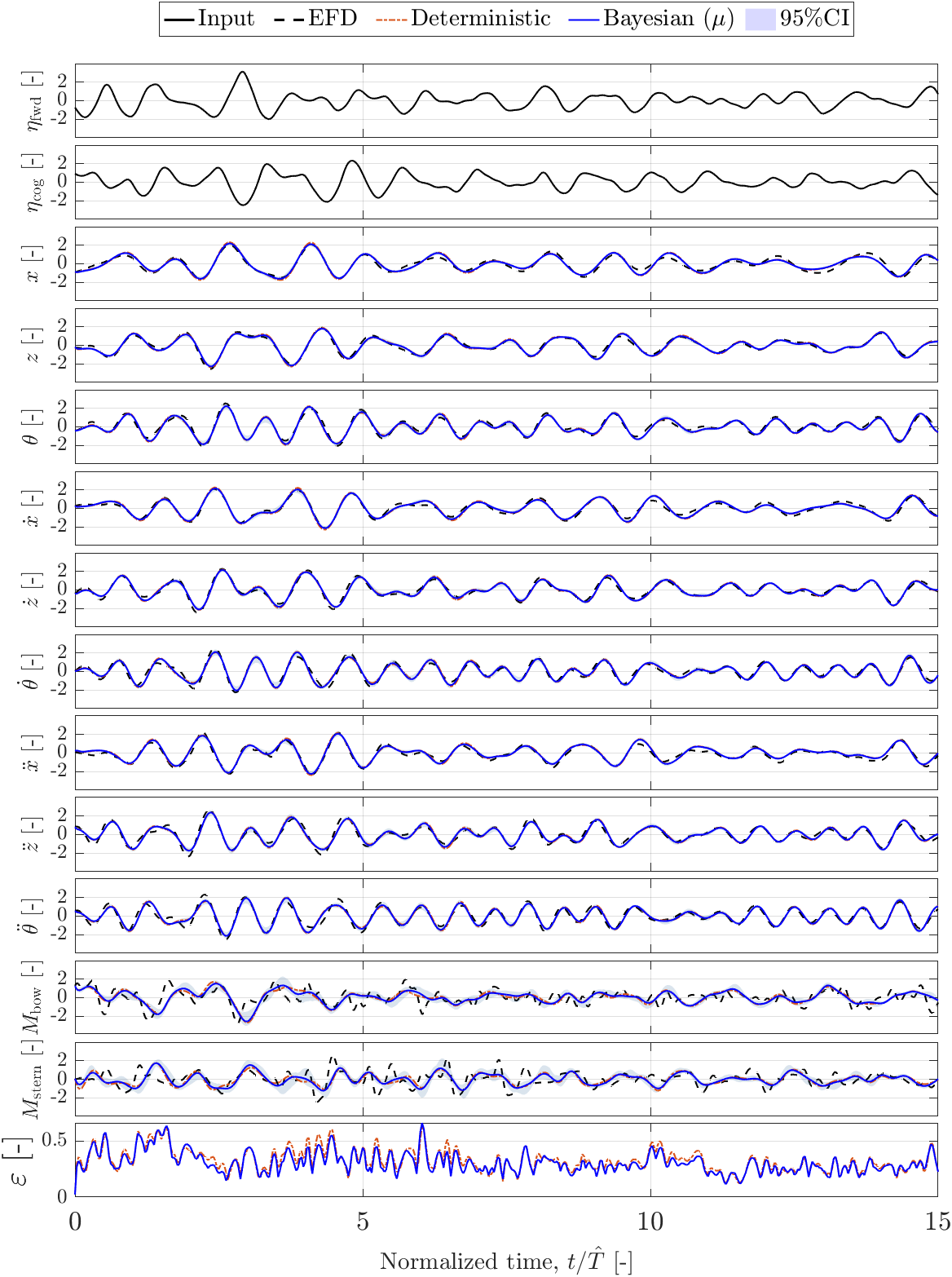}
    \caption{Standardized time series prediction by deterministic and Bayesian Hankel-DMDc. Irregular test wave, selected sequence 1.} \label{fig:irr2irr1}
\end{figure}

\begin{figure}[ht!]
    \centering
    \includegraphics[width=0.9\linewidth]{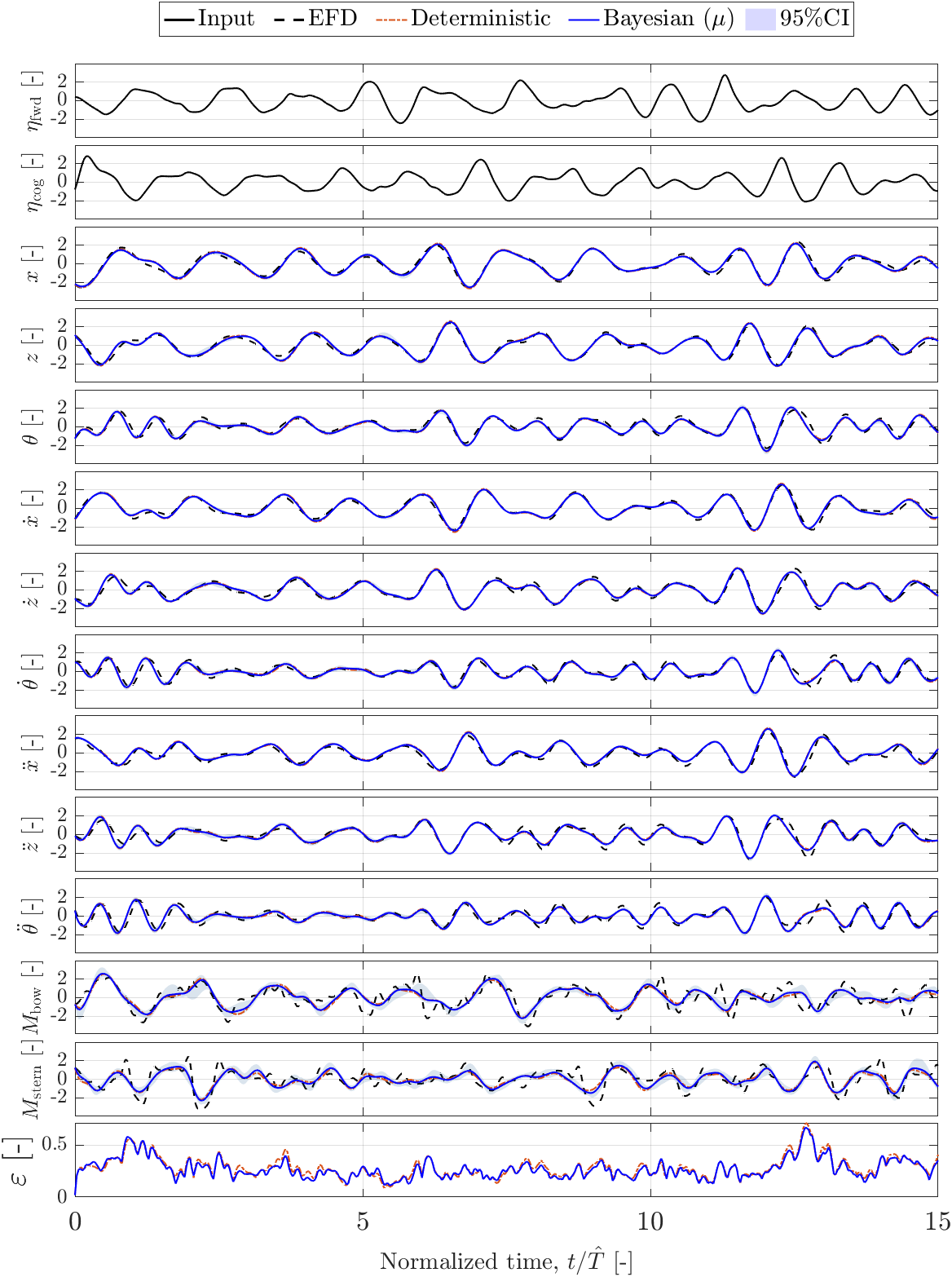}
    \caption{Standardized time series prediction by deterministic and Bayesian Hankel-DMDc. Irregular test wave, selected sequence 2.}\label{fig:irr2irr2}
\end{figure}

Observing the time-resolved predictions in \cref{fig:irr2irr1,fig:irr2irr2} evidences that the great majority of the ANRMSE and NAMMAE errors arise from the prediction of the mooring forces.
This is also confirmed by \cref{tab:jsd_bootstrap} showing that the JSD for the Mbow e Mstern PDFs is an order of magnitude higher than the other variables.
Comparing the box-violin plot in \cref{fig:irr2irr_boxviolin}, it can be noted that the BHDMDc model reduced both ANRMSE and NAMMAE errors, lowering the average error and also reducing the results dispersion. 
The uncertainty in the Bayesian model is very low. This is consistent with the high accuracy achieved for vessel dynamics predictions and indicates the robustness of the model to hyperparameter changes in the identified ranges (confirming the rule of thumbs identified in the literature \cite{palmaFAST2025,palma2025sicatamaran}). However, confidence intervals are not sufficiently extended to cover the ground truth for mooring loads, whose predictions are less accurate and larger uncertainties would have been expected.

The remarkable accuracy obtained for the ship dynamics is also reflected in the MBB analysis, as the PDF from EFD and BHDMDc data are very close and their confidence interval are almost always overlapped. The MBB analysis also confirms the reduced accuracy in the estimation of mooring forces.

The difficulty in obtaining an accurate prediction of the mooring forces is due to a combination of measurement noise and artifacts arising from the physical behavior of the mooring lines.
Specifically, the mooring lines rested at the water surface, alternately emerging or being submerged by waves (particularly for the bow mooring line), which induced sudden variations in tension that do not necessarily reflect the system dynamics. 
In addition, insufficient pretensioning caused the lines to be slack at times, further contributing to spurious oscillations in both the mooring lines and the measured loads.
%Additionally, the lines were not always taut due to insufficient pretensioning. These combined effects induced spurious oscillations in the mooring lines and measured loads.
These factors complicated the learning of the system response by the DMD-based system identification, with a consequent increased error in the prediction of mooring forces.

Nevertheless, the DMD-based models capture the low-frequency content of the loads, as can be seen in \cref{fig:irr2irr1,fig:irr2irr2}, achieving a fair estimation of load peaks. Regularization plays a key role, preventing the identification of spurious and unstable dynamics, to which the method is particularly sensitive in noisy data environments. 
\begin{table}[ht!]
\centering
\caption{EV and 95\% confidence lower bound, upper bound, and interval of JSD of the PDFs of variables in the bootstrapped time series.}\label{tab:jsd_bootstrap}
\begin{tabular}{lcccc}
\multicolumn{5}{l}{JSD$_{\xi}$}\\
Variable & EV & q=0.025 & q=0.975 & U \\
\toprule \\
$x$ & 0.0026 & 0.0010 & 0.0061 & 0.0052 \\
$z$ & 0.0031 & 0.0013 & 0.0061 & 0.0049 \\
$\theta$ & 0.0035 & 0.0017 & 0.0076 & 0.0059 \\
$\dot x$ & 0.0021 & 0.0007 & 0.0063 & 0.0057 \\
$\dot z$ & 0.0036 & 0.0015 & 0.0075 & 0.0060 \\
$\dot \theta$ & 0.0041 & 0.0011 & 0.0106 & 0.0095 \\
$\ddot x$ & 0.0045 & 0.0028 & 0.0073 & 0.0045 \\
$\ddot z$ & 0.0071 & 0.0030 & 0.0140 & 0.0110 \\
$\ddot \theta$ & 0.0086 & 0.0028 & 0.0181 & 0.0153 \\
$M_{\mathrm{bow}}$ & 0.0186 & 0.0099 & 0.0302 & 0.0203 \\
$M_{\mathrm{stern}}$ & 0.0274 & 0.0185 & 0.0387 & 0.0202 \\
\midrule 
avg & 0.0077 & 0.0040 & 0.0139 & 0.0099 \\
\bottomrule 
\end{tabular}
\end{table}

\subsection{Irregular-to-regular}
\begin{figure}[ht!]
    \centering
        \includegraphics[width=0.7\linewidth]{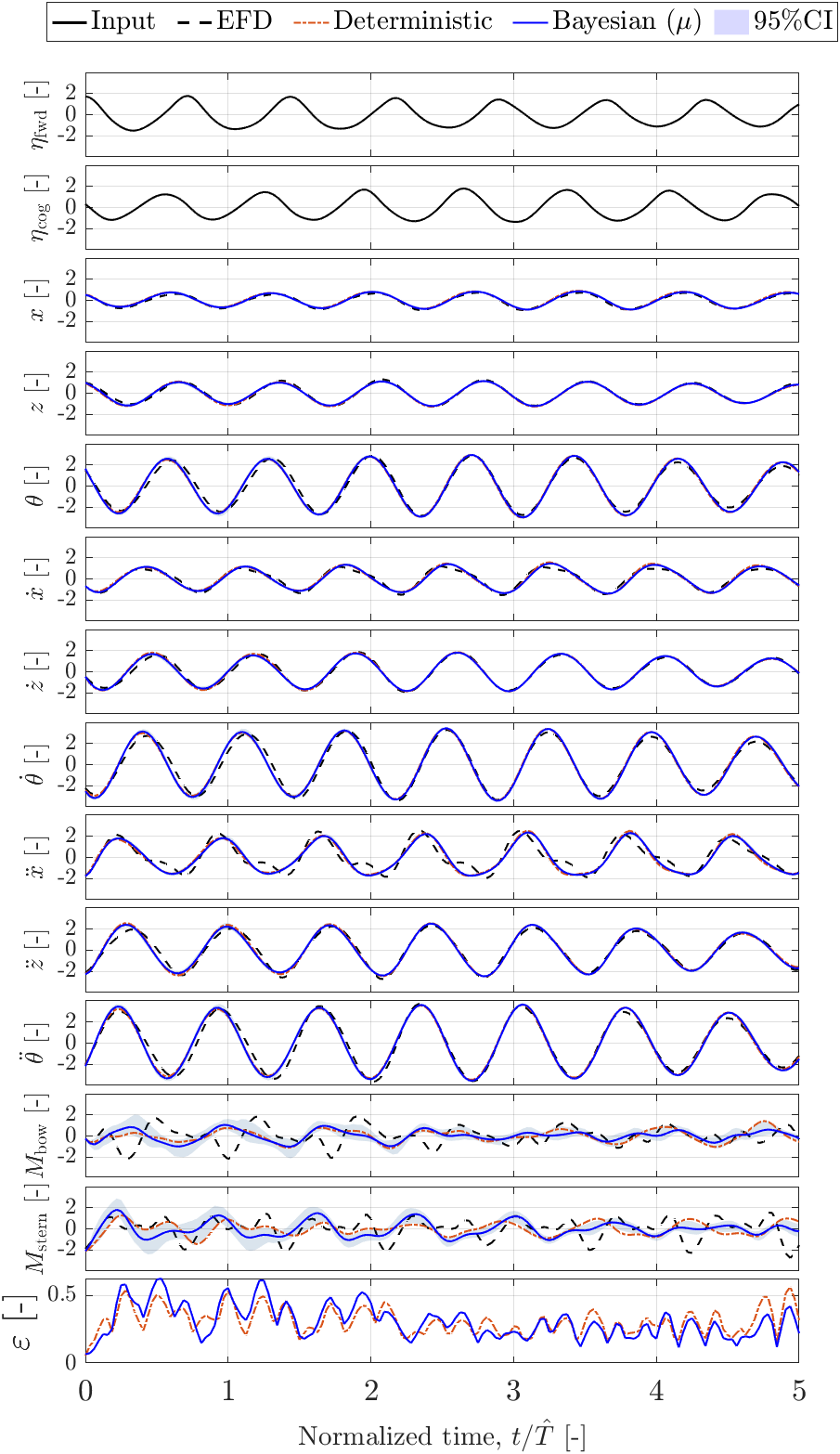}
        \caption{Standardized time series prediction by deterministic and Bayesian Hankel-DMDc. Regular test wave $\lambda_w/L_{m} = 1.5$.}\label{fig:irr2reg1}   
\end{figure}

\begin{figure}[ht!]
    \centering 
        \includegraphics[width=0.7\linewidth]{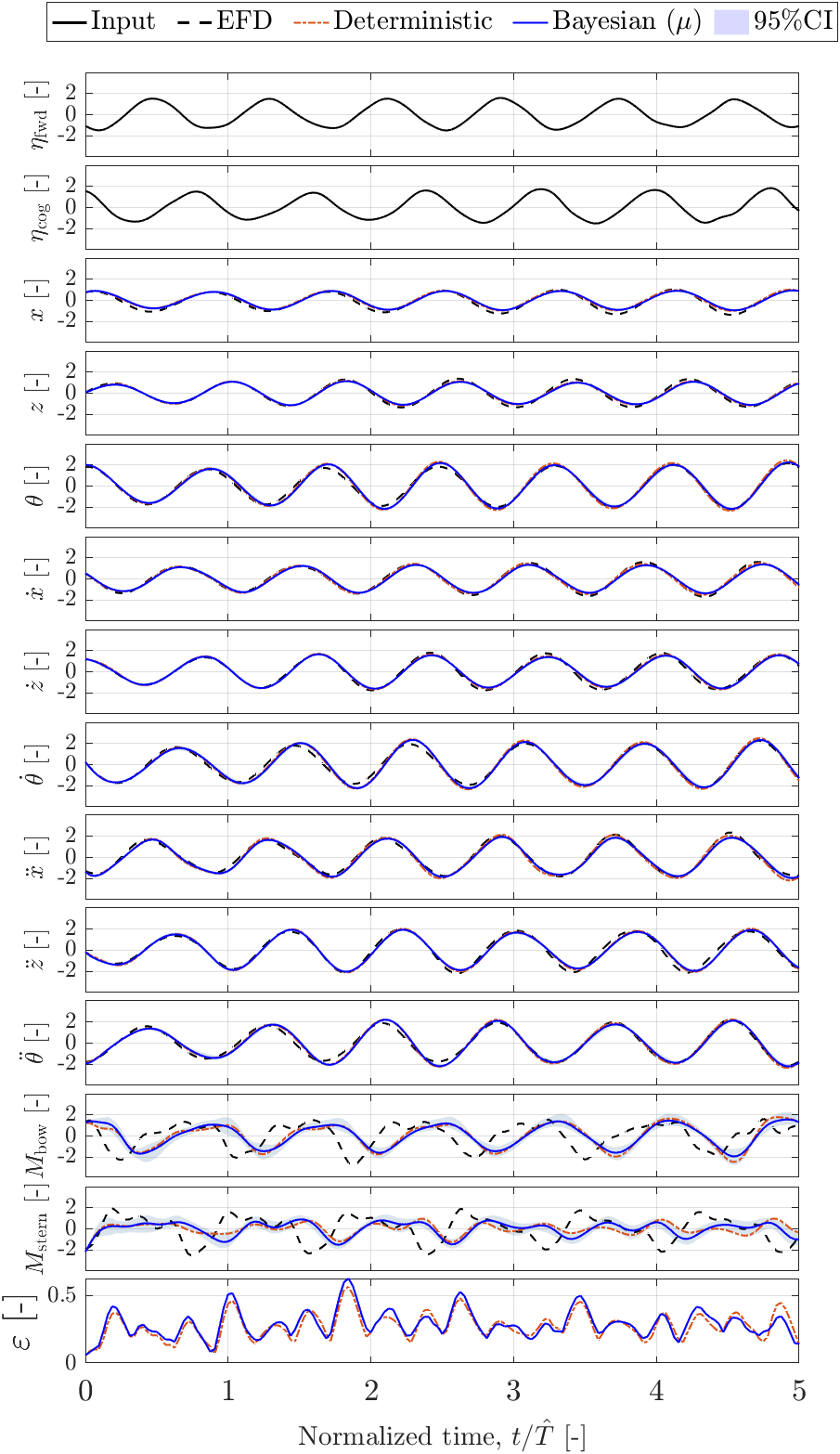}
        \caption{Standardized time series prediction by deterministic and Bayesian Hankel-DMDc. Regular test wave $\lambda_w/L_{m} = 2$.}\label{fig:irr2reg2} 
\end{figure}

\begin{figure}[ht!]
    \centering
        \includegraphics[width=0.7\linewidth]{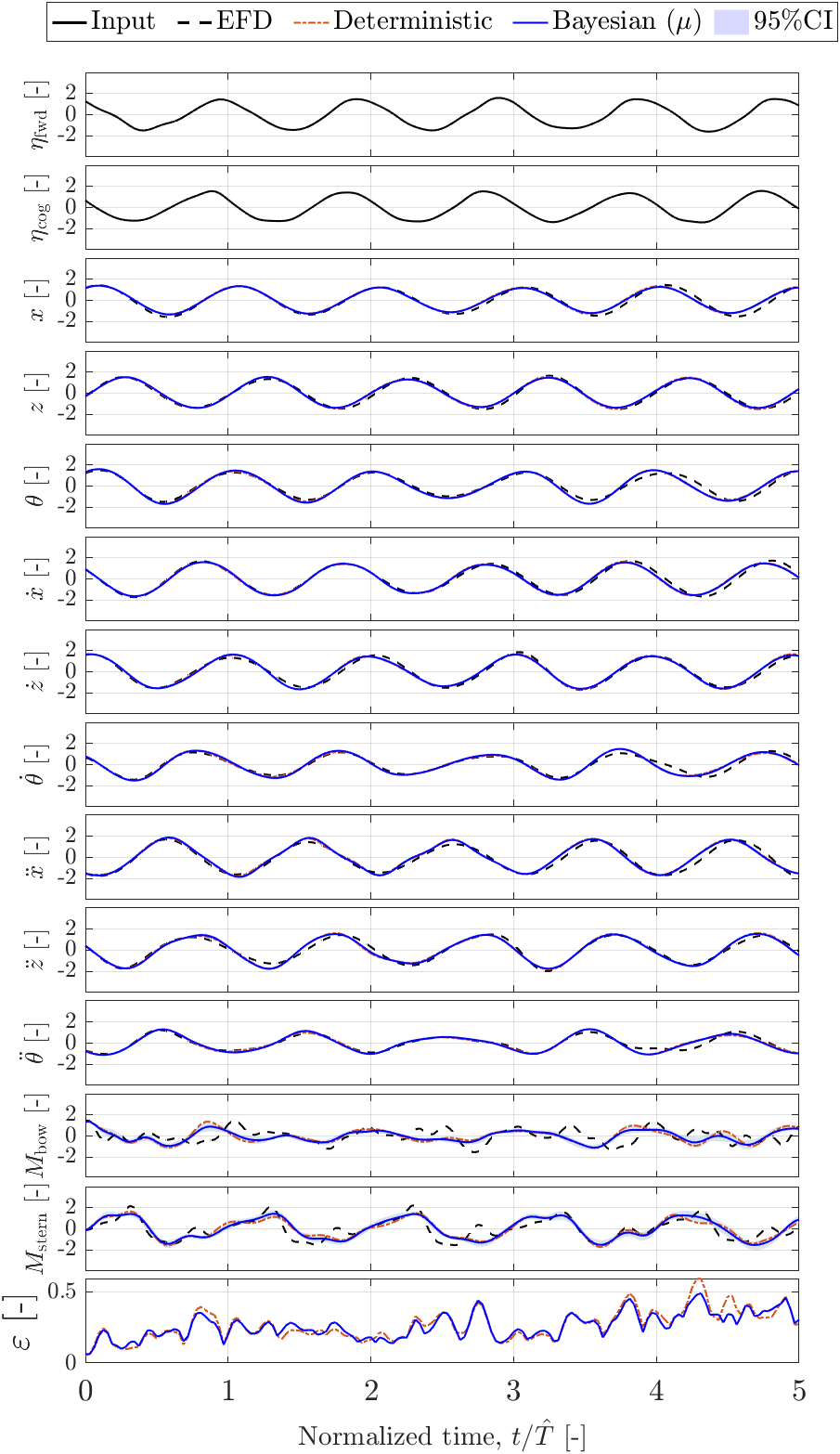}
    \caption{Standardized time series prediction by deterministic and Bayesian Hankel-DMDc. Regular test wave $\lambda_w/L_{m} = 3$.}\label{fig:irr2reg3}  
\end{figure}

This paper first applies the HDMDc and BHDMDc models to predicting ship response in sea conditions different from the training conditions. \Cref{fig:irr2reg1,fig:irr2reg2,fig:irr2reg3} show the results of a selected test sequence for $\lambda_w = 1.5 L_m$, $2 L_m$, and $\lambda_w = 3 L_m$, respectively.
The DMD-based models achieve remarkable accuracy also in these cases for the variables linked to the ship's dynamics, with the largest error being in the reproduction of mooring forces. 
Part of the error clearly arises from an imperfect identification of the load-related subsystem already noted in the irregular waves data results.
In addition, it has been noted that the issue evidenced for the measure of mooring forces was even more prominent in the regular waves case. The model was more effectively pushed downstream during the tests by the regular waves from excitations with shorter wavelength, amplifying slackness of the stern line (loads on the stern mooring are consequently better predicted for higher $\lambda_w$). 
\begin{figure}[ht!]
    \centering
    \captionsetup[subfigure]{justification=centering}    
    \begin{subfigure}[b]{0.49\linewidth} 
        \includegraphics[width=\linewidth]{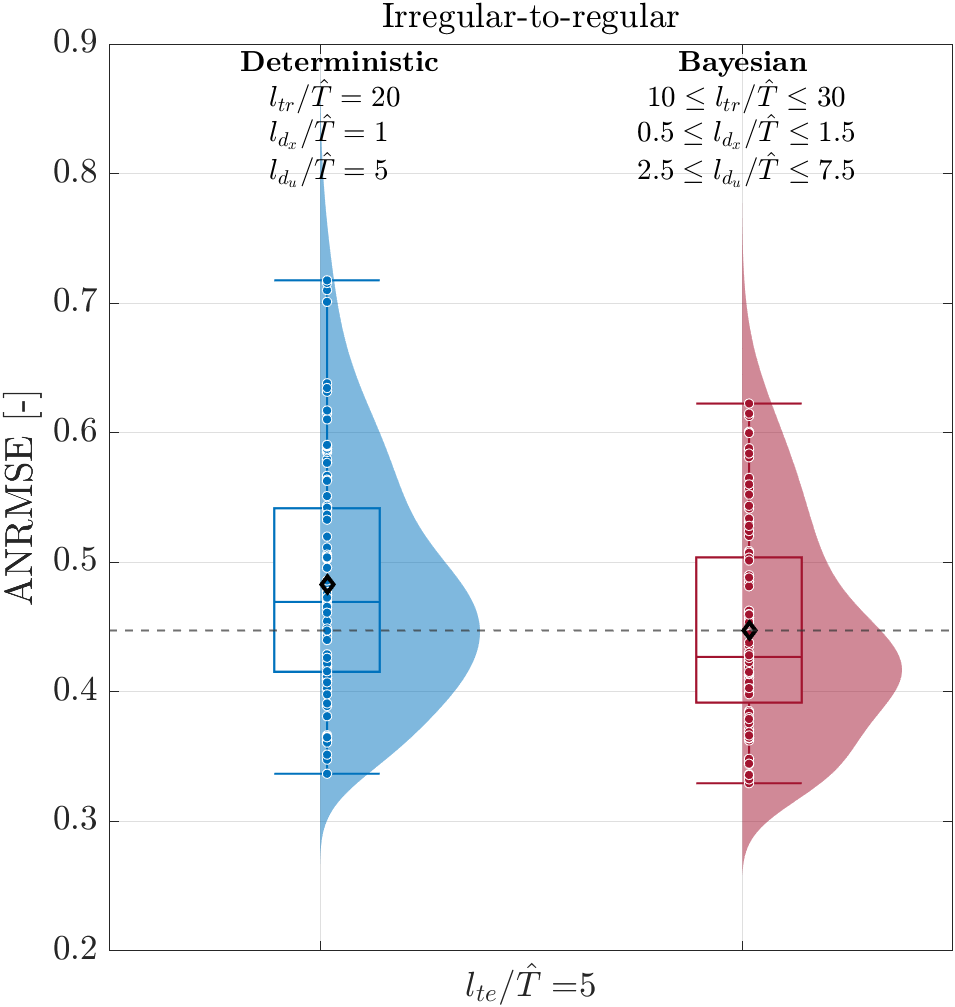}
        \caption{} \label{fig:irr2regANRMSE}
    \end{subfigure}
    \hfill
    \begin{subfigure}[b]{0.49\linewidth} 
        \includegraphics[width=\linewidth]{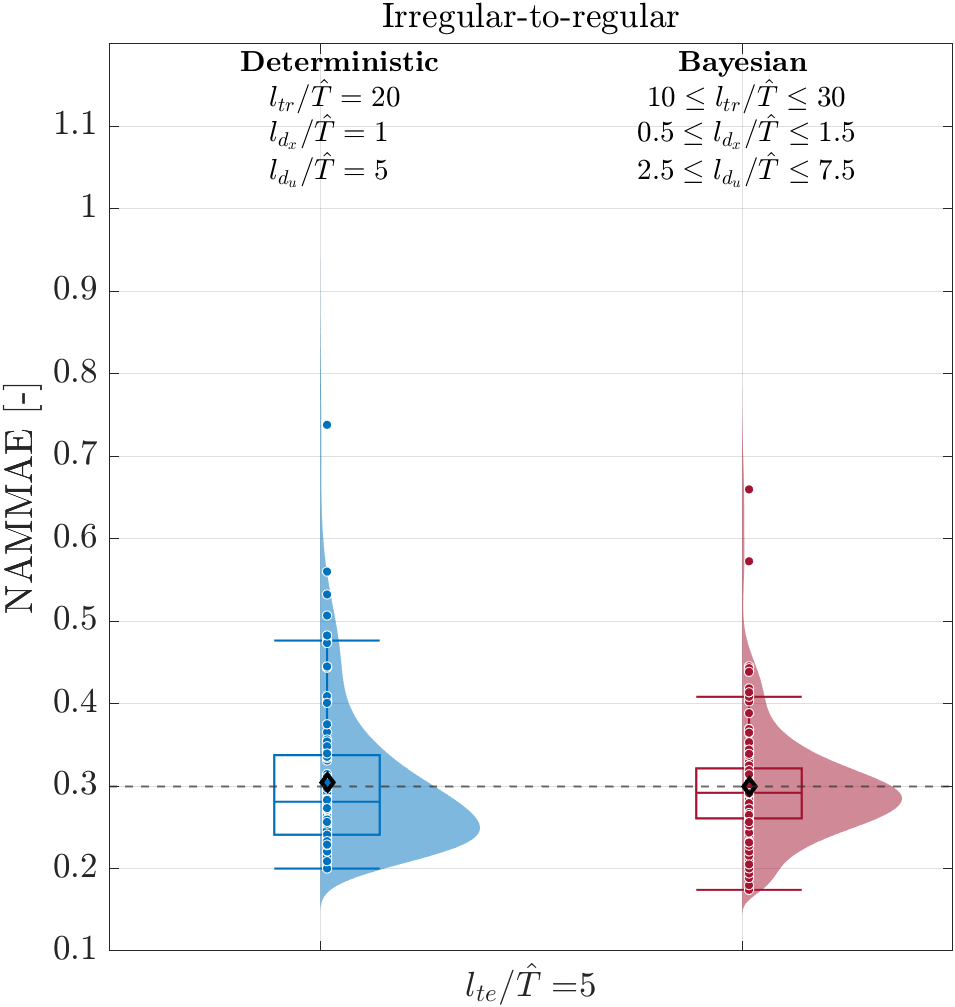}
        \caption{}\label{fig:irr2regNAMMAE}
    \end{subfigure}
    \caption{ANRMSE (\subref{fig:irr2regANRMSE}) and NAMMAE (\subref{fig:irr2regNAMMAE}) box-violin plots. Comparison between deterministic and Bayesian predictions over the regular waves test sequences.}
    \label{fig:irr2reg_boxviolin}
\end{figure}

The box-violin plots containing the statistical analysis of ANRMSE and NAMMAE results are presented in \cref{fig:irr2reg_boxviolin}. It can be noted that the ANRMSE is slightly higher than for irregular-to-irregular wave predictions; however, the NAMMAE is lower. This is coherent with a more pronounced phase error in the prediction of $M_\mathrm{bow}$ and $M_\mathrm{stern}$, while the models still provide a reasonable prediction of the loads' extrema.

The ROMs trained on P-M irregular waves were able to predict vessel responses to regular head waves accurately.
On one side, the regular wave frequencies {\color{black}lie in the decreasing side of} the spectral peak of the training sea state $f({\lambda_w=1.5L_m})=0.86$~Hz, $f({\lambda_w=2L_m})=0.75$~Hz, and $f({\lambda_w=3L_m})=0.61$~Hz, as can be seen from \cref{fig:pmspectrum}.
However, irregular and regular wave excitations have substantial differences in terms of frequency content and phase coherence. 
The successful prediction of regular-wave responses is, hence, an indication that the models have generalized beyond the specific type of excitation seen during training, at least when the new excitation remains within the spectral band covered during training.
This highlights the importance of selecting informative training datasets, \textit{i.e.}, constructing and using training datasets that encompass a sufficiently rich representation of the system dynamics, enabling the development of ROMs capable of generalizing across multiple operating conditions. In this sense, irregular sea states appear promising due to their broadband spectral content, inherently providing a more comprehensive excitation of the system. 

%%%%%%%%%%%%%%%%%%%%%%%%%%%%%%%%%%%%%%%%%%
\FloatBarrier
\section{Conclusions}\label{s:conc}

This work presented the system identification of a small ASV in moored conditions using the HDMDc and its uncertainty-aware extension, BHDMDc. The methods were applied to experimental data collected in the towing tank of CNR-INM under both irregular and regular head-wave conditions. The ASV under investigation features a recessed moon pool, which induces nonlinear responses due to sloshing, thereby increasing the modelling challenge. 

The deterministic HDMDc formulation successfully captured the dominant dynamics of the vessel motions, providing accurate predictions of surge, heave, and pitch across all tested conditions. The prediction of mooring forces was more challenging and exhibited larger errors, mainly due to measurement noise and extremely nonlinear effects related to the partial slackening and intermittent immersion of the mooring lines. Nevertheless, the models were able to reproduce the low-frequency components and general trends of the mooring loads, offering a meaningful approximation of their temporal evolution.

The Bayesian extension introduced uncertainty quantification by propagating the variability of the HDMDc hyperparameters through Monte Carlo sampling. The Bayesian model consistently improved prediction accuracy compared to the deterministic version, also reducing the dispersion of the results as measured by ANRMSE and NAMMAE.

A key outcome of this study is the demonstrated generalization capability of HDMDc and BHDMDc models trained exclusively on irregular-wave data. The trained models were able to accurately predict the ASV response in regular-wave conditions, a different excitation regime although still characterized by frequencies within the spectral range of the training sea state. This result highlights the importance of constructing informative training datasets that encompass a sufficiently rich representation of the system dynamics. In particular, irregular sea states, due to their broadband spectral content, inherently provide a more comprehensive excitation of the system, enabling the development of ROMs capable of generalizing across multiple operating conditions within the same frequency band. 
Further extending the approach to include, e.g., different sea states, wave directions, and operating conditions, would represent a key enabling technology for digital twins of marine systems, where reliable real-time predictions and uncertainty quantification are essential for control, monitoring, and decision support applications. 
In this context, the interpolation of parametric reduced-order models \citep{Farhat2008,Farhat2011} may represent a viable strategy to overcome the intrinsic limitations of using a single, albeit informative, training set, by combining multiple locally trained models to construct a global predictive model that retains accuracy across a broader range of operating conditions.

\section*{Acknowledgements} 
This research was funded by the Italian Ministry of University and Research through the National Recovery and Resilience Plan (PNRR), CN00000023 -- CUP B43C22000440001, “Sustainable Mobility Center” (CNMS), Spoke 3 “Waterways". The authors are also grateful to the US Office of Naval Research for its support in the methodological development through NICOP Grants N62909-21-1-2042 and N62909-24-1-2102.

\section{Conflicts of interest}
Author Lorenzo Minno was employed by the company Codevintec S.r.l. The remaining authors declare that the research was conducted in the absence of any commercial or financial relationships that could be construed as a potential conflict of interest.

\section*{CRediT authorship contribution statement}
\textbf{Giorgio Palma:} Conceptualization, Methodology, Software, Validation, Formal Analysis, Investigation, Data Curation, Writing - Original Draft, Visualization.
\textbf{Ivan Santic:} Investigation, Data Curation, Resources, Writing - Original Draft.
\textbf{Andrea Serani:} Methodology, Resources, Writing - Review \& Editing.
\textbf{Lorenzo Minno:} Resources.
\textbf{Matteo Diez:} Conceptualization, Methodology, Investigation, Resources, Writing - Review \& Editing, Supervision, Project Administration, Funding Acquisition.

All authors have read and agreed to the published version of the manuscript.

% \appendix
% \section{Example Appendix Section}\label{s:app1}


%Bibliography
\bibliographystyle{unsrt}  
\bibliography{biblio}  

\appendix
\end{document}